\documentclass{article}

\usepackage{PRIMEarxiv}

\usepackage[utf8]{inputenc} 
\usepackage[T1]{fontenc}    
\usepackage{hyperref}       
\usepackage{url}            
\usepackage{booktabs}       
\usepackage{amsfonts}       
\usepackage{nicefrac}       
\usepackage{microtype}      
\usepackage{lipsum}
\usepackage{fancyhdr}       
\usepackage{graphicx}       
\graphicspath{{media/}}     
\usepackage{subcaption}
\usepackage{amsmath}
\usepackage{amsfonts}
\usepackage{tikz}
\usepackage{multirow}
\usepackage{algorithm}
\usepackage{algpseudocode}
\usepackage{optidef}
\usepackage{comment}
\usepackage{amssymb}
\usepackage{xcolor}
\usepackage{lscape}
\usepackage{array}
\newtheorem{remark}{Remark}

\newcommand\MyBox[2]{
  \fbox{\lower0.75cm
    \vbox to 1.7cm{\vfil
      \hbox to 1.7cm{\hfil\parbox{1.4cm}{#1\\#2}\hfil}
      \vfil}%
  }%
}

\title{ByteShield: Adversarially Robust End-to-End Malware Detection through Byte Masking}

\author{
  Daniel Gibert\\
  Artificial Intelligence Research Institute \\
  Consejo Superior de Investigaciones Científicas\\
  Spain\\
  \texttt{daniel.gibert@iiia.csic.es}
  \And
  Felip Manya\\
  Artificial Intelligence Research Institute \\
  Consejo Superior de Investigaciones Científicas\\
  Spain\\
  \texttt{felip@iiia.csic.es} \\
}

\begin{document}
\maketitle

\begin{abstract}
Research has proven that end-to-end malware detectors are vulnerable to adversarial attacks. In response, the research community has proposed defenses based on randomized and (de)randomized smoothing. However, these techniques remain susceptible to attacks that insert large adversarial payloads. To address these limitations, we propose a novel defense mechanism designed to harden end-to-end malware detectors by leveraging masking at the byte level. This mechanism operates by generating multiple masked versions of the input file, independently classifying each version, and then applying a threshold-based voting mechanism to produce the final classification. Key to this defense is a deterministic masking strategy that systematically strides a mask across the entire input file. Unlike randomized smoothing defenses, which randomly mask or delete bytes, this structured approach ensures coverage of the file over successive versions. In the best-case scenario, this strategy fully occludes the adversarial payload, effectively neutralizing its influence on the model’s decision. In the worst-case scenario, it partially occludes the adversarial payload, reducing its impact on the model's predictions. By occluding the adversarial payload in one or more masked versions, this defense ensures that some input versions remain representative of the file's original intent, allowing the voting mechanism to suppress the influence of the adversarial payload. Results achieved on the EMBER and BODMAS datasets demonstrate the suitability of our defense, outperforming randomized and (de)randomized smoothing defenses against adversarial examples generated with a wide range of functionality-preserving manipulations while maintaining high accuracy on clean examples.
\end{abstract}

\section{Introduction}
Malware detection is the task of identifying malicious software, i.e computer programs designed to damage or exploit computer systems. Malware detection has evolved significantly over time as both malware and detection technologies have advanced. Recently, machine learning (ML) and deep learning (DL) have been widely adopted as an additional layer of defense against malware. Traditional machine learning approaches~\cite{10.1145/2857705.2857713,2018arXiv180404637A,harang2020sorel20m,GIBERT2022117957} rely on manually engineered features (e.g., byte N-grams, API calls, strings, etcetera) to build classifiers for distinguishing between legitimate and malicious files. Deep learning or end-to-end approaches~\cite{DBLP:conf/aaai/RaffBSBCN18,DBLP:conf/iclr/KrcalSBJ18,GIBERT2021102159,DBLP:conf/aaai/RaffFZAFM21}, on the other hand, leverage the power of neural networks to automatically learn complex patterns from raw data. These approaches have shown promise in detecting zero-day malware, with the academic community and anti-malware companies claiming very high detection rates~\cite{10102612,2018arXiv180404637A,DBLP:conf/aaai/RaffBSBCN18}.

Although promising, the use of machine learning and deep learning for detecting malware is not a panacea. These detectors are vulnerable to adversarial examples, i.e., malicious executables that have been deliberately manipulated to evade detection. More specifically, end-to-end detectors~\cite{DBLP:conf/aaai/RaffBSBCN18} can be evaded by patching unused bytes and injecting adversarial payloads of hundreds~\cite{DBLP:journals/corr/abs-1901-03583,10.1145/3473039} or thousands of bytes~\cite{8553214,10319738,YUSTE2022102643,9437194,gibert2024certifiedadversarialrobustnessmachine}. This is because these classifiers learn correlations between specific byte patterns and benign or malicious behavior and artifacts. Attackers exploit these associations by injecting or modifying bytes in non-functional code regions, thereby misleading the detector into misclassifying a malicious file as benign.

Various defenses have been proposed to improve the robustness of end-to-end malware detectors against adversarial examples, including adversarial training~\cite{287238}, randomized smoothing-based methods~\cite{gibert2023_randomizedsmoothing,huang2023rsdel}, and (de)randomized smoothing methods~\cite{10.1145/3605764.3623914,gibert2024certifiedadversarialrobustnessmachine,saha2024drsm}. More specifically, a recent method~\cite{10.1145/3605764.3623914,gibert2024certifiedadversarialrobustnessmachine,saha2024drsm} based on (de)randomized smoothing has shown promise in defending against adversarial examples. This method provides formal robustness guarantees against patch and content insertion attacks. Nevertheless, this defense, which is based on (1) splitting a file into chunks, (2) independently classifying each chunk, and (3) performing a majority voting to determine the final prediction, 
comes with a trade-off: a decrease in accuracy on clean examples. Furthermore, this defense is vulnerable against large adversarial payloads. The larger the adversarial payload, the larger the proportion of chunks that will be influenced, thereby breaking the majority voting mechanism and bypassing detection. 

\begin{figure}[ht]
    \includegraphics[width=0.6\textwidth]{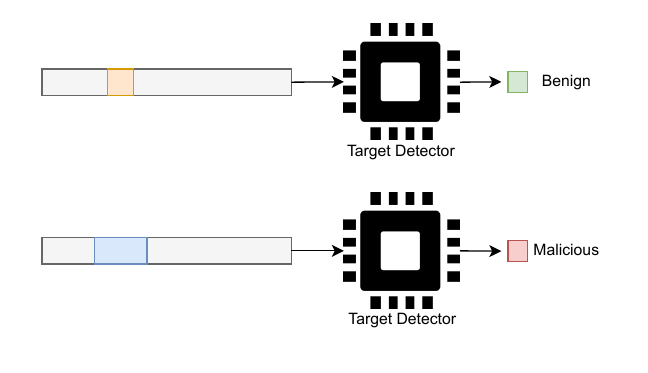}
    \centering
    \caption{Illustration of the masking process. In the top part, an adversarial payload (highlighted in orange) is injected within a malicious executable file. When this adversarial example is processed by the malware detector, the adversarial payload causes the detector to incorrectly classify the malicious executable as benign. In the bottom part, a mask (highlighted in blue) covers the region containing the adversarial payload. By masking the adversarial payload, its influence on the detector is neutralized, thereby correctly classifying the executable as malicious.}
    \label{fig:masking_example}
\end{figure}

In this paper, we advance the state of the art by proposing a novel robust mechanism that addresses the limitations of (de)randomized smoothing,  i.e., low clean accuracy and vulnerability against large payloads. The core idea behind the proposed approach is to mask the adversarial payload that the attacker has injected within the executable to bypass the malware detector. By masking the payload, as illustrated in Figure~\ref{fig:masking_example}, the adversarial attack will be neutralized, i.e., the attacker's manipulations to the executable are completely blocked. In consequence, the classifier's prediction will not be influenced by the attacker, allowing the classifier to correctly classify the executable as malicious. 

In practice, however, the location and size of the adversarial payload are unknown, complicating the masking process. To address this challenge, we propose a sequential masking strategy that systematically strides a fixed-size mask across the entire input file using a sliding window approach, generating a set of masked versions of the input file in which different portions of the file are selectively occluded. By sliding the mask across the entire executable, our strategy guarantees complete coverage of the input space. This ensures that at least one masked version of the input sample will have the adversarial payload fully occluded, provided that the mask size is larger than the payload and that the payload resides in a single contiguous region. Under these conditions, our masking strategy effectively neutralizes the adversarial influence regardless of where the injection occurs.

Following the masking process, the predictions for each masked version are aggregated using a threshold-based decision logic, where an executable file is classified as malware if at least $T$ versions are labeled as malicious. The rationale underlying this approach is that when the masked region overlaps with the adversarial payload, the classifier's prediction should revert to the correct label, as the influence of the adversarial payload is eliminated. Conversely, if the masked region does not cover the payload, the classifier may produce incorrect predictions. However, a consistent minority of correct predictions across multiple masked versions still indicate the presence of an adversarial payload. For clean executables, i.e., executables that have not been manipulated to evade detection, we expect the classifier to consistently predict the correct class.

To assess the suitability of our masking strategy, we conducted a series of experiments using the MalConv architecture~\cite{DBLP:conf/aaai/RaffBSBCN18}, a well-established end-to-end model for malware detection, and the EMBER~\cite{2018arXiv180404637A} and the BODMAS~\cite{bodmas} datasets, two widely used public benchmarks for malware detection. Results demonstrate that ByteShield outperforms competing methods based on  randomized~\cite{huang2023rsdel} and (de)randomized smoothing~\cite{gibert2024certifiedadversarialrobustnessmachine,saha2024drsm}, achieving comparable f1-score and setting a new benchmark in adversarial accuracy. Additionally, our approach offers significantly faster inference time compared to existing randomized smoothing defenses, making it both robust and computationally efficient (see Section \ref{sec:computational_time}).

The rest of the paper is organized as follows. Section \ref{sec:related_work} describes the related work. Section \ref{sec:threat_model} describes the threat model considered in this work. Section~\ref{sec:methodology} presents the main components of ByteShield. Section~\ref{sec:evaluation} compares the proposed defense with randomized and (de)randomized smoothing defenses. Lastly, Section~\ref{sec:conclusions} summarizes the conclusions of our research and outlines future lines of research.

\section{Related Work}
\label{sec:related_work}
This Section reviews relevant literature in end-to-end malware detection, adversarial attacks, and defenses.
\subsection{End-to-End Malware Detection}
\label{sec:end_to_end_detection}
End-to-end malware detection~\cite{DBLP:conf/aaai/RaffBSBCN18,DBLP:conf/iclr/KrcalSBJ18,GIBERT2021102159,DBLP:conf/aaai/RaffFZAFM21} involves training a machine learning model to identify malicious software directly from raw data, such as binary files represented as a sequence of bytes. In contrast to traditional methods~\cite{10.1145/2857705.2857713,2018arXiv180404637A}, which rely on feature engineering to extract hand-crafted features based on expert knowledge, end-to-end detection learns patterns directly from the data during training, reducing dependency on domain-specific knowledge.

Executable files can range in size from a few kilobytes to several megabytes, resulting in sequences that may span millions of timesteps, where each byte represents a single time step. As processing these large sequences requires substantial computational resources, researchers have proposed using shallow convolutional neural networks (CNNs)~\cite{DBLP:conf/aaai/RaffBSBCN18,DBLP:conf/iclr/KrcalSBJ18,DBLP:conf/aaai/RaffFZAFM21}. These networks consist of a small number of convolutional layers with large kernels and strides, followed by a global max-pooling or average-pooling layer.\footnote{Despite its shallow architecture, the original MalConv model~\cite{DBLP:conf/aaai/RaffBSBCN18} was trained on a DGX-1 server, well beyond the means of the majority of research groups.}  Despite their adoption and success for malware detection, end-to-end models have been shown to be vulnerable to adversarial attacks.

\subsection{Functionality-Preserving Adversarial Attacks}
\label{sec:adv_attacks}
\begin{figure}[ht]
    \includegraphics[width=0.5\textwidth]{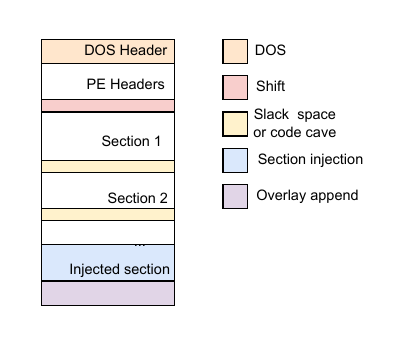}
    \centering
    \caption{Illustration of a PE file structure and regions targeted by the adversarial attacks.}
    \label{fig:pe_structure}
\end{figure}

Numerous attacks have been proposed to craft adversarial examples that bypass end-to-end malware detectors without breaking the original functionality of the executables. As illustrated in Figure~\ref{fig:pe_structure}, these attacks either patch unused bytes~\cite{8844597,DBLP:journals/corr/abs-1901-03583,10.1145/3473039} or inject adversarial content~\cite{8553214,10.1145/3473039,YUSTE2022102643,9437194,gibert2024certifiedadversarialrobustnessmachine}. Below, we describe several notable attacks. 

Kolosnjaji et al.~\cite{8553214} proposed a gradient-based attack that manipulates a malicious file by appending a set of carefully selected bytes at the end of the file, a region also known as overlay. Suciu et al.~\cite{8844597} improved upon this approach by optimizing the slack bytes, unused bytes that may exist between sections of the file created by alignment requirements, which can be freely modified without altering the behavior of the computer program.

Demetrio et al.~\cite{DBLP:journals/corr/abs-1901-03583} used feature attribution to identify the most influential features contributing to malware classification. They found that convolutional neural networks (CNNs), such as MalConv~\cite{DBLP:conf/aaai/RaffBSBCN18}, tend to focus on characteristics of the file headers instead of the data and text sections. Based on this analysis, they proposed an attack algorithm that generates adversarial examples by only changing a few bytes in the DOS header. Similarly, Zhan et al.~\cite{10319738} presented MalPatch, an attack that manipulates the bytes in both the DOS header and the overlay~\cite{8553214} of executables to inject the adversarial payload.

Demetrio et al.~\cite{10.1145/3473039} presented three attacks, targeting different regions of the PE file: (1) Full DOS, which modifies all bytes in the DOS header; (2) Extend, which injects the payload into the space created by extending the DOS header; and (3) Shift, which injects the payload between the headers and the first section of the file.

Yuste et al.~\cite{YUSTE2022102643} developed an attack that generates adversarial examples by dynamically creating code caves (unused blocks) between sections in malware binaries. Then, a genetic algorithm optimizes the adversarial payload inserted within these caves.

Demetrio et al.~\cite{9437194} introduced GAMMA, a functionality-preserving attack that relies on the injection of benign content at the end of the file, or within newly-created sections. Then, a genetic algorithm minimizes the amount of benign content that needs to be injected to bypass the target malware detector.

\subsection{Adversarial Defenses}
\label{sec:adv_defenses}
Following we detail the methods that have been proposed to defend against the adversarial attacks listed in Section~\ref{sec:adv_attacks}.

Lucas et al.~\cite{287238} investigated the effectiveness of adversarial training for end-to-end malware detection. In their work, they used three attacks to train the classifier: (1) in-place replacement attack (IPR)~\cite{10.1145/3433210.3453086}, (2) displacement attack (Disp)~\cite{10.1145/3433210.3453086} and (3) padding attack~\cite{8844597}. \footnote{We excluded the in-place replacement~\cite{10.1145/3433210.3453086} and displacement attacks~\cite{10.1145/3433210.3453086} from our attack evaluation due to their dependency on a commercial IDA Pro license, which was not available for this study.} Their findings revealed that while adversarial training can improve robustness against known attacks, i.e., attacks included in the training data, it fails to deter novel, unknown attacks. Furthermore, adversarial training presents significant limitations: (1) it is computationally expensive due to the need of generating adversarial examples during training; (2) it requires prior knowledge of potential attack types; and (3) the model's performance on clean, non-adversarial samples decreases. \footnote{Due to these limitations, we also excluded adversarial training from our defense evaluation.} 

Recently, various methods have been proposed based on randomized smoothing~\cite{gibert2023_randomizedsmoothing,huang2023rsdel}, a defense mechanism designed to improve robustness of a model against adversarial attacks by adding random noise to the input data. Randomized smoothing works as follows: (1) During training, a base classifier is trained based on input data that has been slightly modified by adding random noise; (2) At inference time, multiple noisy versions are generated and independently classified by the classifier. Afterwards, a majority voting mechanism is applied to determine the final prediction. Building on this concept, Gibert et al.~\cite{gibert2023_randomizedsmoothing} proposed to randomly mask bytes while Huang et al.~\cite{huang2023rsdel} proposed to randomly delete bytes. Both approaches, while they are slightly more robust against adversarial attacks with low budgets, do not provide sufficient protection and can  be easily bypassed by strong content-injection attacks.

Gibert et al.~\cite{10.1145/3605764.3623914,gibert2024certifiedadversarialrobustnessmachine} and Saha et al.~\cite{saha2024drsm} adapted (de)randomized smoothing for malware detection. In their work, a base classifier is trained to make classifications based on a subset or chunk of bytes. At test time, a given input example is split into sequential, non-overlapping chunks of fixed size, and each chunk is independently classified. Then, majority voting is applied to determine the final classification. This defense achieves greater robustness against the attacks listed in Section~\ref{sec:adv_attacks}. However, it suffers from a significant drop in clean accuracy. This reduction is due to the model's reliance on small chunks during classification, which limits the amount of information available in each chunk, leading to less accurate predictions on clean, non-manipulated examples. Furthermore, this defense can be bypassed by injecting an adversarial payload equal to the size of the original malicious executable as it will result in having at least 50\% of the chunks classified as benign. 

\section{Threat Model}
\label{sec:threat_model}
This section defines the threat model considered in this work, including the attack scenario, as well as the capabilities of both the attacker and defender.

\subsection{Attack Scenario}
The goal of a malicious actor is to craft a malicious executable that evades detection by the end-to-end detector. To this end, the attacker can freely manipulate the binary in a way that its original functionality is preserved while misleading the model into classifying it as benign.

Let $x =\left[x_{0}, x_{1}, ..., x_{L} \right] \in \mathcal{X} \subseteq \{0, 1, ..., 255 \}^{L}$ be the original malicious executable, represented as a sequence of raw bytes, $\mathcal{T} (x)\subseteq \mathcal{X}$ be the set of functionality-preserving transformations of $x$, 
and $f: \mathcal{X} \to \left[ 0, 1 \right] $ be the end-to-end detector, which returns a malicious score:
\begin{itemize}
    \item If $f(x) \ge 0.5$, $x$ is classified as malicious.
    \item If $f(x) < 0.5$, $x$ is classified as benign.
\end{itemize}
The goal of the malicious actor is to find a variant $x' \in \mathcal{T} (x)$ such that:
$$ f(x') < 0.5 \; and \;x'\equiv x$$
Then, the objective of the attacker is to solve the following optimization problem:
\begin{align}
\min_{x' \in \mathcal{T}_R(x)} \quad & f(x') \label{eq:attack-objective-distributed} \\
\text{subject to} \quad & f(x) \geq 0.5, \label{eq:malicious-input-distributed} \\
                        & x' \equiv x, \label{eq:functional-constraint-distributed} \\
                        & \sum_{r=1}^{R} \|x'_{\mathcal{R}_r} - x_{\mathcal{R}_r}\|_0 \leq B, \label{eq:global-budget-constraint} \\
                        & \mathcal{R}_1(x), \dots, \mathcal{R}_R(x) \text{ are disjoint.} \label{eq:disjoint-constraint}
\end{align}

where:
\begin{itemize}
    \item \( \mathcal{T}_R(x) \) denotes the set of functionally equivalent transformations applied across \( R \) disjoint regions.
    \item \( x_{\mathcal{R}_r} \) denotes the subvector of \( x \) indexed by region \( \mathcal{R}_r \).
    \item \( B \) is the total modification budget across all regions.
    \item \( x' \equiv x \) enforces that functionality is preserved.
    \item \( f(x') \) is the classifier’s predicted maliciousness score for the adversarial input \( x' \).
\end{itemize}

\subsection{Attacker Capabilities}
The attacker can have either white-box or black-box access to the target malware detector, each defining different levels of adversarial capacity:
\begin{itemize}
    \item In the white-box setting, the attacker has full access to the end-to-end model's architecture, parameters, and gradients. This allows the attacker to (1) perform gradient-based optimization to craft the adversarial payloads, and (2) identify salient or influential byte positions that contribute heavily to the model's decision.
    \item In the black-box setting, the attacker does not have access to the internal architecture or parameters of the detector. Instead, the attacker may query the model and observe its output score $f(x')$ or the final label.
\end{itemize}

Regardless of the access level, the attacker has full control over the injected bytes, which are typically placed in non-functional or appendable regions of the executable. This ensures that the resulting adversarial example $x'$ maintains the same functionality as $x$, i.e., $x' \equiv x$.\footnote{While optimization in white-box settings is significantly more efficient due to having access to the gradients to compute the most effective perturbation direction to alter the model's prediction, it is not a realistic threat in practice as malicious actors typically do not have access to the internals of the model. }

\subsection{Defender Capabilities}
The defender operates a fixed end-to-end detection system $f: \mathcal{X} \to \left[ 0, 1 \right]$, which takes raw byte sequences as input and outputs a maliciousness score. The defender is not aware of whether an input has been adversarially manipulated and has no knowledge of the location, size, or structure of any adversarial payload. More specifically, we assume the defender:
\begin{itemize}
    \item Has no access to adversarial examples during training.
    \item Cannot assume prior knowledge of the attacker's strategy, capabilities, or access level (white-box or black-box).
    \item Can process inputs at inference time and may apply transformations, but does not modify the underlying model architecture or parameters.
\end{itemize}
The goal of the defender is to create a robust model by design, without requiring knowledge of the attacker's manipulations.

\section{ByteShield Defense}
\label{sec:methodology}
In this paper, we introduce ByteShield, a novel byte masking strategy that effectively limits the ability of the adversarial payloads to alter the outcome of end-to-end malware detectors by design. The proposed byte masking strategy relies on the fact that if we can figure out where the adversarial payload is located and, thereby, mask it, its influence on the detector will be neutralized and the detector will correctly classify the file as malicious.

In practice, however, the location and size of the adversarial payload are unknown. Moreover, the attacker may distribute the payload across multiple locations of the executable. Due to the global max-pooling and average-pooling layers commonly used in end-to-end architectures for malware detection on raw byte sequences, the impact of the injected payload on the model's prediction is largely independent of its position within the file. To address this challenge, we propose a byte masking scheme that sequentially masks regions of the input file and analyzes the classifier's responses. Here is how our defense works:
\begin{enumerate}
    \item Given an input file, we generate multiple versions in which parts of the executable are occluded following the masking strategy 
  outlined in Section~\ref{sec:sequential_masking_strategy}.
    \item Each masked version is then independently classified, and the resulting classifier's predictions are analyzed to determine the final classification.
\end{enumerate}

For the masked versions in which the adversarial payload is fully obscured, the classifier is expected to make accurate predictions that match the true label. If the adversarial payload is only partially obscured, it may still influence the classifier's prediction, though not necessarily enough to flip the label from malicious to benign. Under attack, most predictions are expected to deviate from the true label. However, there is likely at least one instance where the adversarial payload is completely occluded, and the classifier's prediction will agree with the true label. In contrast, when no attack is present, all predictions are expected to agree with the true label.

Our defense leverages this pattern, where a minority of predictions consistently yield the correct label when under attack, to detect the presence of an adversarial payload. In the specific context of malware detection, where there are only two classes (benign and malicious) it is illogical to inject an adversarial payload into a benign executable so as it is detected as malicious. Consequently, if there is a disagreement between predictions it can safely be assumed that the executable is malicious.

Unfortunately, classifying a file as malware if only one masked version is malicious can result in a large number of false positives, where benign executables are incorrectly classified as malicious. To address this issue, we adopt a less conservative decision rule by classifying an executable as malicious if at least $T$ masked versions are labeled as malicious. This threshold-based approach strikes a balance between reducing false positives and maintaining robust detection of adversarial examples.

\begin{figure*}[ht]
    \includegraphics[width=\textwidth]{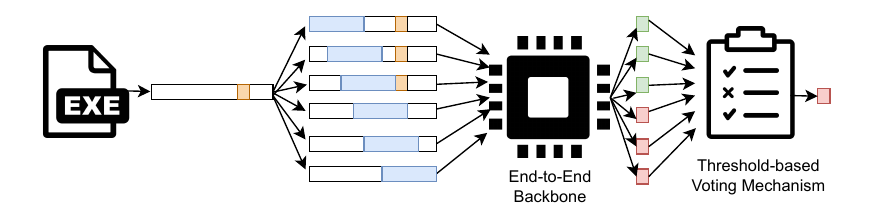}
    \centering
    \caption{Illustration of our byte masking scheme using a stride equals to 10\% and a mask size equals to 50\% (blue squares) of the original file's size. Given an input example (containing an adversarial payload represented with the orange square), our defense sequentially strides the mask through different locations, and independently classifies each masked version with the base classifier. Finally, the threshold-based voting mechanism determines the final classification by aggregating the individual predictions for each masked version of the input example.}
    \label{fig:sequential_masking_strategy}
\end{figure*}

\subsection{Masking Strategy}
\label{sec:sequential_masking_strategy}
The sequential masking strategy, illustrated in Figure~\ref{fig:sequential_masking_strategy}, operates by systematically occluding different regions of the executable using a sliding window approach. By sliding the mask across the entire executable, the potential adversarial payload will be eventually occluded, regardless of its location or size (provided that the mask size is larger than the payload). 

While the core masking operation remains consistent across training and inference, its intent and method of application differ.

\paragraph{Training Phase}
A key component of our defense is enabling a classifier $f: \mathcal{X} \rightarrow [0, 1]$ to learn robust representations from partially occluded inputs, ensuring it can make accurate predictions even when parts of the executable are masked. To achieve this, we train an end-to-end malware classifier to make predictions on masked versions of the training samples, as defined in Algorithm~\ref{alg:masking_training}. Given a training example $x \in \mathcal{X} $, we generate a masked input $\tilde{x}$ by applying the masking operation
$$\tilde{x} = \text{MASK}(x, idx, m)$$
where $idx$ represents the starting position in the file at which the mask is applied, and $m$ specifies the size of the mask in bytes. The classifier is then trained to minimize the loss $\mathcal{L}(f(\tilde{x}, y))$
 where $y\in\{0,1\}$
 is the true label of the unmasked input $x$.
 
 The $\text{MASK}(x, idx, m)$ function, described in Algorithm~\ref{alg:masking_function}, transforms an input byte sequence by masking the bytes from position $idx$ to position $  idx + m - 1$. Since executable files vary in size, the mask size is defined as a percentage~$M \in \mathbb{Z}$ of the file's length $L$, where $1 \leq M < 100$. The size in bytes $m$, is then computed relative to the length of the input example as follows: 
$$m = \lceil L \times \frac{M}{100} \rceil \;\; M \in  \mathbb{Z}, \;1 \leq M < 100$$
This ensures that the masking operation is applied consistently across executables of varying lengths.

\begin{algorithm}[ht]
    \caption{ByteShield's training procedure.}\label{alg:masking_training}%
    \begin{algorithmic}
        \Require training dataset $D_{train}$, end-to-end detector $f$ with parameters $\theta$, mask size $M$, batch size $B$, optimizer $opt$, loss function $criterion$
        \State \text{Initialize parameters}$\: \theta $
        \For{i = 0 \textbf{to}  MAX\_EPOCHS}
            \State Shuffle $D_{train}$
            \For{each mini-batch $(X, Y) \subset D_{train}$ with batch size $B$}
                \State $\tilde{X} \gets []$ \Comment{Initialize list of masked examples}
                \For{$b=0$ \textbf{to} $B-1$} \Comment{Process batch of examples one by one}
                    \State $x \gets X[b]$ \Comment{Get a single example x}
                    \State $L \gets length(x)$ \Comment{Get size of x}
                    \State $m = \lceil L \times \frac{M}{100} \rceil$
                    \State $idx \gets \text{RandomInteger}(0, L - m)$ \Comment{Get starting position of the mask}
                    \State $\tilde{x} \gets \text{MASK}(x, idx, m)$  \Comment{Apply mask of size m to example x starting at position idx}
                    \State Append $\tilde{x}$ to $\tilde{X}$ 
                \EndFor
        
                \State $\tilde{Y} \gets f(\tilde{X}, \theta)$ \Comment{Forward pass on masked data}
                \State $Loss \gets criterion(Y, \tilde{Y})$ \Comment{Compute loss}
                \State Compute gradients for Loss with respect to $\theta$
                \State Use the optimizer to update $\theta$ based on the computed gradients
                \EndFor    
            \EndFor
    \end{algorithmic}
\end{algorithm}

\begin{remark}
    To mask subsequences of bytes, we use the "PAD" token, which is typically used to ensure that all sequences have the same size and is represented as a vector of zeros. This vector is not updated during training, ensuring that the masked positions do not contribute to the classifier's output. For instance, if the size of the embedding layer is $E=8$, then the embedding of the 'PAD' token is $[0,0,0,0,0,0,0,0]$. Note that while we could have defined an additional 'MASK' token, its effect would be functionally equivalent to that of the 'PAD' token.   
\end{remark}

\begin{algorithm}
    \caption{Implementation of the MASK function.}\label{alg:masking_function}
    \begin{algorithmic}
        \Require Input sequence $x$, starting position $idx$, mask size $m$
        \State $\tilde{x} \gets x$ \Comment{Create copy of $x$}
        \For{$j = idx$ \textbf{to} $idx + m - 1$} 
        \State $\tilde{x}[j] \gets 256$ \Comment{Replace the byte at position j with 256 ("PAD" token)}
        \EndFor
        \State \Return $\tilde{x}$ 
        
    \end{algorithmic}
\end{algorithm}

\paragraph{Inference Phase}
At inference time, a sliding window approach is used to systematically occlude the entire input space. This procedure is outlined in Algorithm~\ref{alg:sequential_masking_prediction_function}. Let $x$ be an input executable of length $L$, and let $f$ be the base classifier. The masking strategy at inference time is parametrized by:
\begin{itemize}
    \item $M \in \mathbb{Z}$, $1 \le M < 100 $, the mask size expressed as a percentage of the file's length.
    \item $S \in \mathbb{Z}$, $1 \le S < M $, the stride size expressed as a percentage of the file's length. We restrict $S$ to be smaller than the mask size $M$ to allow for overlapping masks, which ensure full coverage of the input file and increase the likelihood that the adversarial payload is fully occluded in at least one masked version.
\end{itemize}
The mask size in bytes is computed as:
$$m = \lceil L \times \frac{M}{100} \rceil \;\;  \text{where}\;\; M \in  \mathbb{Z}, \;1 \leq M < 100$$
and the stride size in bytes is given by:
$$s = \lceil L \times \frac{S}{100} \rceil\;\;  \text{where}\;\; S \in  \mathbb{Z}, \;1 \leq S < M$$
Then, at inference time, we generate a set of masked versions ${\tilde{x}^{(1)}, \tilde{x}^{(2)}, ..., \tilde{x}^{(N)}}$, where each masked input is defined by:
$$ \tilde{x}^{(n)} =  \text{MASK}(x, idx, m)\;\; \text{with} \;\; idx = n \times s$$
where the number of masked versions $N$ is computed as:
$$N = \lceil \frac{L - m}{s} \rceil$$
This ensures that the mask strides through the input in steps of size $s$, generating non-overlapping or partially overlapping masked versions of the input $x$ depending on the values of $M$ and $S$.

\begin{remark}
Striding the mask at regularly spaced positions, i.e., skipping $s$ bytes between each application, is key to reducing the computational cost of the defense, as it significantly limits the number of masked versions that must be generated and evaluated during inference. Without this optimization, applying the mask at every byte position would require generating $L - m$ masked versions, which is computationally infeasible for large executables. For example, a file with size $L=1,000,000$ with a mask size $m = 500,000$ would require $500,000$ masked versions, making the method impractical at scale. To address this, we reduce the number of masked versions by defining the stride as a percentage of the file size. For instance, with $M=50\%$ and $S=1\%$, only  50 masked versions are created, significantly reducing the computational overhead while still maintaining effective coverage of the file.
\end{remark}

Afterwards, each masked example $\tilde{x}^{(n)}$ is independently classified by $f$, producing a score:
$$f(\tilde{x}^{(n)}) \in \left[0.0, 1.0\right]$$
We then assign a binary label to the masked example as follows:
\begin{itemize}
    \item $\tilde{y}^{(n)} = 1 \quad \text{if} \quad f(\tilde{x}^{(n)}) \geq  0.5$
    \item $\tilde{y}^{(n)} = 0 \quad \text{if} \quad f(\tilde{x}^{(n)}) < 0.5$
\end{itemize}

Next, the number of times the classifier assigns each class is counted. Let:
\begin{itemize}
    \item $\text{num\_malicious} =  \sum_{n=0}^{N} \mathbb{I}(\tilde{y}^{(n)} = 1) $ be the number of masked examples classified as malicious.
    \item $\text{num\_benign} =  \sum_{n=0}^{N} \mathbb{I}(\tilde{y}^{(n)} = 0) $ be the number of masked examples classified as benign.
\end{itemize}
The final classification is determined by the following threshold-based voting mechanism:
\begin{itemize}
    \item $\text{If} \; \text{num\_malicious} \geq  T \;\text{then}\; \text{classify} \; x\; \text{as} \;\text{malicious}$
    \item $\text{If} \; \text{num\_malicious} < T  \;\text{then}\; \text{classify} \; x\; \text{as} \; \text{benign}$
\end{itemize}

\begin{algorithm}
    \caption{ByteShield's prediction function.}\label{alg:sequential_masking_prediction_function}
    \begin{algorithmic}
        \Require Input file $x$, mask size $M$, base classifier $f$, threshold $T$, stride $S$
        \State $\text{num\_malicious} \gets 0$
        \State $\text{num\_benign} \gets 0$
        \State $ L \gets length(x)$
        \State $m \gets \lceil L \times \frac{M}{100} \rceil \text{(mask size in bytes)}$
        \State $s \gets \lceil L \times \frac{S}{100} \rceil \text{(stride size in bytes)}$
        \State $N \gets \lceil \frac{L -m }{s} \rceil \text{(number of masked versions)}$
        \For{$n=0 \; \textbf{to}\; N $}
        \State $idx \gets n \times s$ 
        \State $\tilde{x}^{(n)} \gets \text{MASK}(x, idx, m)$ 
        \State $\tilde{y}^{(n)} \gets f(\tilde{x}^{(n)})$ 
        \If{$\tilde{y}^{(n)} \geq  0.5$}
        \State $\text{num\_malicious} \gets \text{num\_malicious} + 1$
        \Else
        \State $\text{num\_benign} \gets \text{num\_benign} + 1$
        \EndIf
        \EndFor
        
        \If{$\text{num\_malicious} \geq T$}
        \State \Return ``Malicious''
        \Else
        \State \Return ``Benign''
        \EndIf
    \end{algorithmic}
\end{algorithm}

\subsubsection{Security Evaluation}
\label{sec:sequential_masking_security_evaluation}
The proposed defense is secure by design. The fundamental principle is that if the adversarial payload is fully masked, the adversary cannot influence the classifier's decision. For certification purposes, we assume access to unlimited computational power and apply the mask at every byte position. In this scenario, as long as the mask size $m$ is greater than the size of the adversarial payload $p$, we can guarantee that the adversarial payload will be masked in at least one instance. This prevents it from influencing the classifier's decision and ensures that at least one prediction will differ from the majority. 
	
Suppose our defense, using a mask size $m$, receives as input a file $x$ with length $L$ that has been altered by injecting an adversarial payload of size $p$, where $p \leq m$. The mask is applied at every position $i$ in $x$, where $i=0,1,2,..., L-m$.
At each position, the mask occludes $m$ bytes of $x$ with the 'PAD' token. Let's define the prediction set as:
$$ F = \{ f(\tilde{x}^{(i)}) | i=0,1,..., L-m \}$$

Given that $m\geq p$, for any position $j$ where the payload starts in $x$, the payload spans $j, j+1, j+2, j+p-1$. Then, any mask applied at position $j$ will cover the payload because 
$$ \text{Masked region} = \{i | i \in [j, j+m-1]\}$$ includes the entire payload region.

Let $f(\tilde{x}^{(j)})$ denote the classifier's prediction for input $x$ with the mask applied starting at position $j$ and let $f(\tilde{x}^{(i)})$ denote the classifier's prediction for input $x$ at any other position where $i \neq j$ (payload partially or not masked).

If the adversarial payload is completely masked, it cannot influence the classifier's decision.  If the payload is partially masked or not masked, it still can influence the classifier $f$. As a result, for some $j$ (payload fully masked) and any $i \neq j$ (payload partially masked or not masked):
$$ f(\tilde{x}^{(j)})  \neq  f(\tilde{x}^{(i)})$$

Since $f(\tilde{x}^{(j)})$ exists in $F$ and diverges from any other prediction $f(\tilde{x}^{(i)})$, our defense can still detect when a file has been manipulated.

Unfortunately, we cannot apply the mask at every position during inference as it is computationally infeasible. Table \ref{tab:number_masked_version_per_stride} illustrates the number of masked versions that would be required for a file of $1,000,000$ bytes in size, various $s$ values, and a fixed mask size $m=500,000$. As the stride size decreases, the number of masked versions increases, quickly leading to prohibitive computational costs.
\begin{table}[ht]
\centering
\caption{Number of masked versions $\textbf{N}$ generated for a $\textbf{1,000,000}$ executable file using a fixed mask size $\textbf{M=50\% (m=500,000)}$ and varying stride values.}
\label{tab:number_masked_version_per_stride}
\begin{tabular}{c|c|c|c|c|c}
\toprule
\textbf{L}         & \textbf{M}  & \textbf{S}      & \textbf{m}       & \textbf{s}      & \textbf{N}       \\ \midrule
1,000,000 & 50 & 0.0001 & 500,000 & 1      & 500,000 \\
1,000,000 & 50 & 0.001  & 500,000 & 10     & 50,000  \\
1,000,000 & 50 & 0.01   & 500,000 & 100    & 5,000   \\
1,000,000 & 50 & 0.1    & 500,000 & 1,000  & 500     \\
1,000,000 & 50 & 1      & 500,000 & 10,000 & 50      \\
1,000,000 & 50 & 2      & 500,000 & 20,000 & 25      \\
1,000,000 & 50 & 5      & 500,000 & 50,000 & 10     \\ \bottomrule
\end{tabular}%
\end{table}

Furthermore, the attacker is not limited to the injection of a single adversarial payload but may distribute the payload across multiple regions of the executable file. For example, techniques such as the code caves~\cite{YUSTE2022102643} and gamma attacks~\cite{9437194} introduce space between sections or create new sections within the executable, allowing the adversarial payload to be split across these newly-created spaces. This fragmented injection limits our ability to formally certify the accuracy of our defense, as we may be unable to guarantee the complete occlusion of all payload fragments. Consequently, if any influential fragment remains unmasked, the adversarial payload may still be able to flip the classifier's decision. However, in practice, not all fragments contribute the same towards the model's final decision, i.e., some regions may be more influential than others depending on their content. As a result, occluding a subset of the adversarial fragments may be sufficient to neutralize the attack's impact and correctly classify the file. This observation is supported by the empirical results presented in Section~\ref{sec:adversarial_evaluation}, where partially occluding the payloads is often enough to recover the correct label. While formal certification remains challenging due to both the fragmented nature of the state-of-the-art (SOTA)  attacks and the use of striding in our masking strategy, our defense demonstrates strong empirical robustness against SOTA attacks, outperforming competing defenses. 

\section{Evaluation}
\label{sec:evaluation}
In this section, we evaluate the effectiveness of our defense. Sections \ref{sec:datasets} and \ref{sec:end_to_end_detectors} present the datasets and end-to-end malware detectors used in our experiments, respectively. Sections \ref{sec:masked_training_analysis} and \ref{sec:hyperparameter_tuning_analysis} analyze the impact of masked training, including the effects of mask size, stride size, and threshold, on the performance of our defense. Sections \ref{sec:clean_evaluation} and \ref{sec:adversarial_evaluation} evaluate the performance of our defense under clean and adversarial conditions, respectively. Section \ref{sec:temporal_robustness} evaluates their performance under concept drift. Finally, Section \ref{sec:computational_time} reports the computational cost of training and inference for the classifiers.

\subsection{Datasets}
\label{sec:datasets}
To evaluate our defense, we use two datasets: (1) the EMBER dataset and (2) the BODMAS dataset. 
\subsubsection{EMBER Dataset}
The EMBER dataset \cite{2018arXiv180404637A} consists of 400,000 benign and 400,000 malicious executables, all timestamped prior to 2018. The EMBER dataset contains data collected before and during 2017. Consequently, the EMBER dataset is used to train our models and assess their robustness against adversarial attacks. To train, validate, and test our models, we randomly split the data into training (80\%), validation (10\%), and test (10\%) sets.\footnote{We will provide the hashes of the samples for both the EMBER and BODMAS datasets to enable other researchers to reproduce our work.} 

\subsubsection{BODMAS Dataset}
The BODMAS dataset \cite{bodmas} consists of 77,142 benign and 57,293 malicious timestamped executables collected from both 2019 and 2020, with family information (581 families). This dataset is used to evaluate the performance of ByteShield under concept drift.

\subsection{End-to-End Malware Detectors}
\label{sec:end_to_end_detectors}
\begin{figure}[ht]
    \includegraphics[width=0.7\columnwidth]{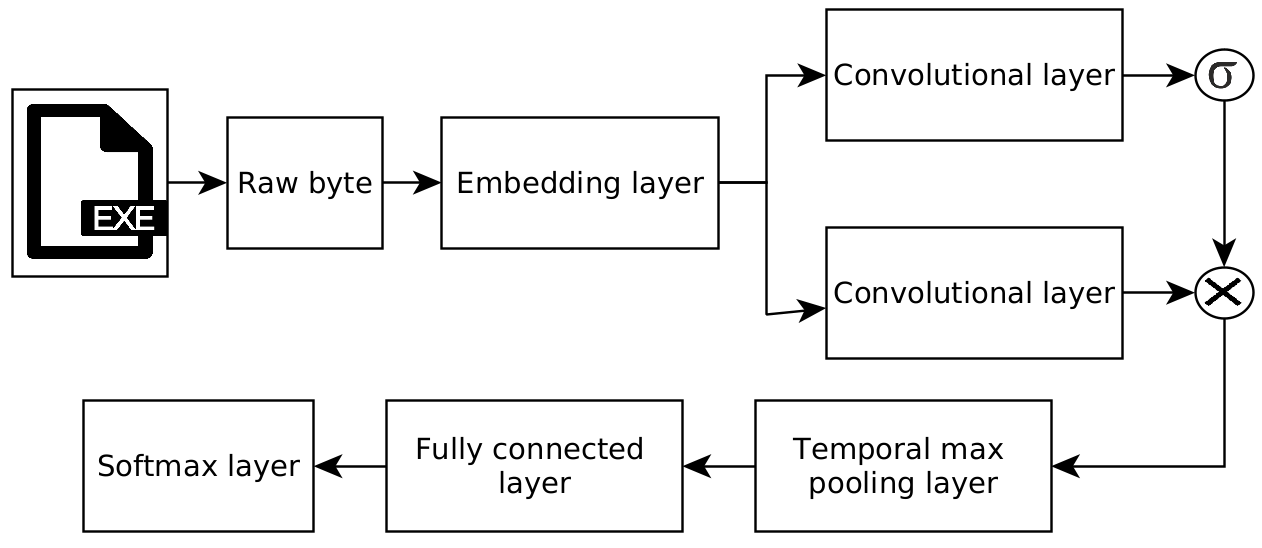}
    \centering
    \caption{Graphical depiction of MalConv's architecture.}
    \label{fig:malconv}
\end{figure}

We selected the MalConv architecture~\cite{DBLP:conf/aaai/RaffBSBCN18} as the backbone for our experiments because it is not only a well-established end-to-end architecture, but also the primary target of the adversarial attacks developed to date~\cite{8844597,DBLP:journals/corr/abs-1901-03583,10.1145/3473039,8553214,YUSTE2022102643,9437194,gibert2024certifiedadversarialrobustnessmachine}. Given its central role in the literature, using MalConv is the natural and obvious choice for evaluating the effectiveness of our defense.\footnote{A recent study~\cite{gibert2024certifiedadversarialrobustnessmachine} has shown that other end-to-end architectures are also susceptible to the attacks developed against MalConv, reinforcing the relevance of using MalConv as a representative benchmark. Nevertheless, additional experiments involving other architectures are provided in the Appendix.} The MalConv architecture, depicted in Figure~\ref{fig:malconv}, consists of an embedding layer, a gated convolutional layer, followed by global max-pooling and a feed-forward layer. 

Using MalConv as the backbone, we trained the following detectors:
\begin{itemize}
    \item  \textit{MalConv}. This detector is used as the undefended baseline.
    \item \textit{RsDel-MalConv}. This detector implements the randomized deletion scheme~\cite{huang2023rsdel} with a 97\% probability of deleting a byte.
    \item \textit{DRS-MalConv}. This detector implements the (de)randomized smoothing scheme~\cite{10.1145/3605764.3623914,saha2024drsm}, where each input example is split into contiguous, non-overlapping chunks. In our experiments, we split the executables into $k$ chunks, where $k \in \{5, 10, 20, 50\}$
    \item \textit{ByteShield-MalConv}. This detector implements our sequential masking strategy using MalConv as base classifier, with various mask sizes $M\in \{10, 20, 30, 40, 50\}$, strides $S\in \{1, 2, 5\}$, and classification thresholds $T\in \{1,2,..., 15\}$. The selected values of $M$ are designed to evaluate the robustness of the defense against attacks that inject varying amounts of adversarial content. A small mask equals to 10\% of the executable's content serves as a lower bound. A mask size $M=10\%$ has minimal impact on the input but it may be insufficient to occlude large adversarial payloads. A mask size $M=30\%$ offers a balanced trade-off between occluding a meaningful portion of the input and preserving the classifier's performance. In contrast, a mask size $M=50\%$ is intended to test the limits of our defense under a more aggressive masking strategy by maximizing occlusion.
\end{itemize}

\subsection{Effects of Masked Training}
\label{sec:masked_training_analysis}
Masked training, detailed in Algorithm \ref{alg:masking_training}, is a critical component of our end-to-end malware detection system. 
Table \ref{tab:masked_training} reports the accuracy, true positive rate (TPR), false positive rate (FPR), and f1-score of various models (with stride $S=1$ and threshold $T=1$) trained with and without masking on the test set. The results for the detectors without masked training correspond to applying our masking strategy only at inference time, using the baseline (undefended) MalConv as backbone. 

\begin{table}[ht]
\centering
\caption{Effects of masked training on detection performance across different mask sizes.}
\label{tab:masked_training}
\begin{tabular}{l|l|cccc}
\toprule
                                         & \textbf{Detectors}                                       & \textbf{Accuracy} & \textbf{TPR} & \textbf{FPR} & \textbf{F1-Score} \\ \midrule
\multirow{5}{*}{\shortstack{Without \\ masked \\training}} & ByteShield-MalConv(M=10)                   &  95.51  & 0.9925  & 0.0877  & 0.9593  \\
                                         & ByteShield-MalConv(M=20)                   &  94.57  & 0.9934  & 0.1089  & 0.9512  \\
                                         & ByteShield-MalConv(M=30)                   &  93.86  & 0.9940  & 0.1248  & 0.9452  \\
                                         & ByteShield-MalConv(M=40)                   &  93.23  & 0.9943  & 0.1385  & 0.9400  \\
                                         & ByteShield-MalConv(M=50)                   &  92.57  & 0.9942  & 0.1524  & 0.9345 \\ \midrule
\multirow{5}{*}{\shortstack{With\\ masked\\ training}}   & ByteShield-MalConv(M=10)                   & 98.12  & 0.9849  & 0.0231  & 0.9824\\
                                         & ByteShield-MalConv(M=20)                   & 97.97  & 0.9791  & 0.0197  & 0.9809\\
                                         & ByteShield-MalConv(M=30)                   & 98.12  & 0.9849  & 0.0230  & 0.9825\\
                                         & ByteShield-MalConv(M=40)                   & 98.27  & 0.9807  & 0.0150  & 0.9837\\
                                         & ByteShield-MalConv(M=50)                   & 98.16  & 0.9863  & 0.0239  & 0.9823\\ \bottomrule   
\end{tabular}%
\end{table}
Across all metrics (e.g., accuracy, FPR, and f1-score) with the exception of TPR, models trained with masked inputs consistently outperform those trained without. This indicates that models trained without masked training detect malware effectively but are prone to misclassifying benign samples as malicious. This is because these models only see uncorrupted byte sequences during training. Therefore, the models learn to rely heavily on specific byte-level patterns. However, during inference, portions of the byte sequence containing discriminative features may be masked, preventing the model from observing the patterns it learned during training. In addition, in the absence of masked training, the performance of the models degrade slightly as the mask size increases. 
For example, the f1-score drops from $0.9593$ ($M=10$) to $0.9345$ ($M=50$). This degradation in performance for models trained without masking is expected: as the mask size increases, a larger portion of the byte sequence becomes occluded, yet these models are conditioned to operate on fully intact inputs. In contrast, this drop is not significant for the models trained with masked training, where performance remains largely consistent across different mask sizes. This stability can be attributed to their exposure to incomplete data during training, which encourages the model to generalize better when portions of the input are masked. 

\subsection{Hyperameter Tuning}
\label{sec:hyperparameter_tuning_analysis}

\begin{figure*}[htbp]
  \centering
  \captionsetup[subfloat]{font=scriptsize} 

  \subfloat[TPR vs FPR (M=10)\label{fig:2}]{
    \includegraphics[width=0.3\textwidth]{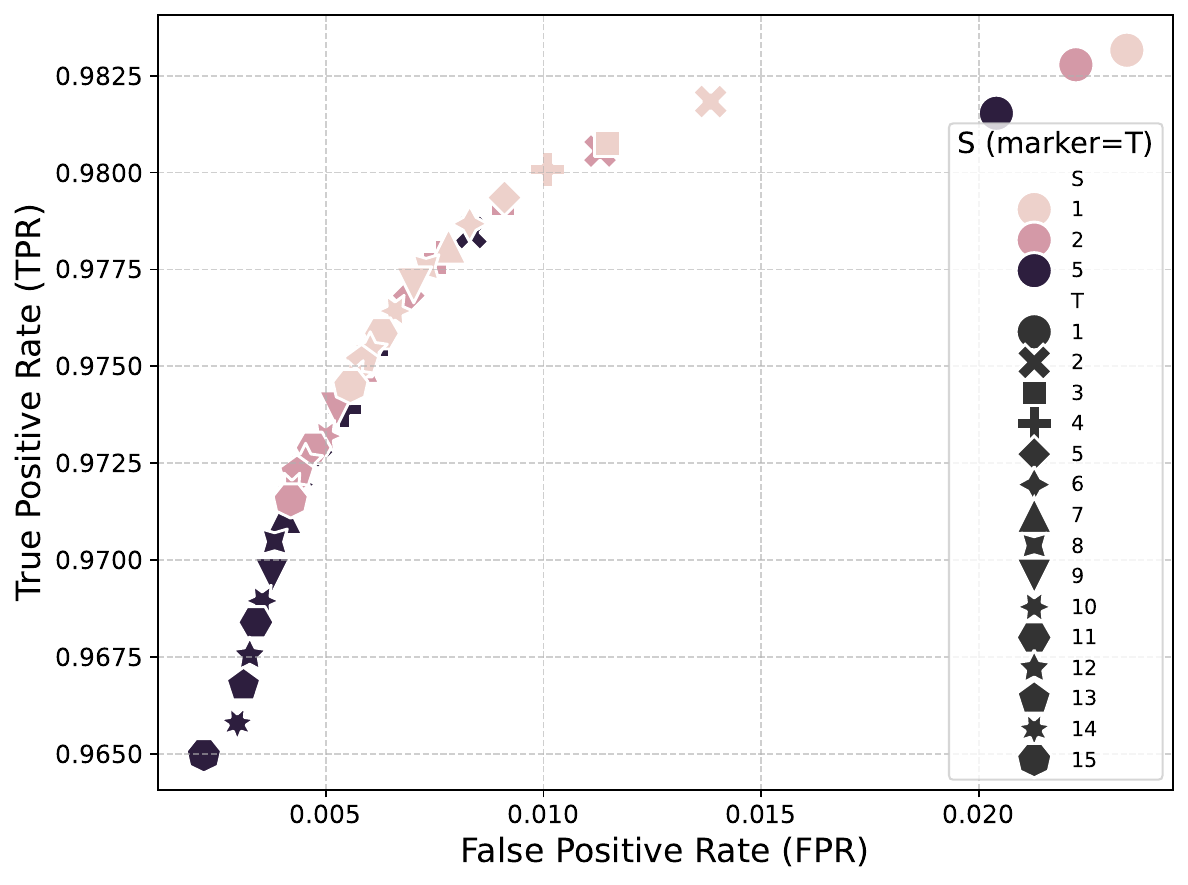}
  }\hfill
  \subfloat[TPR vs FPR (M=20)\label{fig:3}]{
    \includegraphics[width=0.3\textwidth]{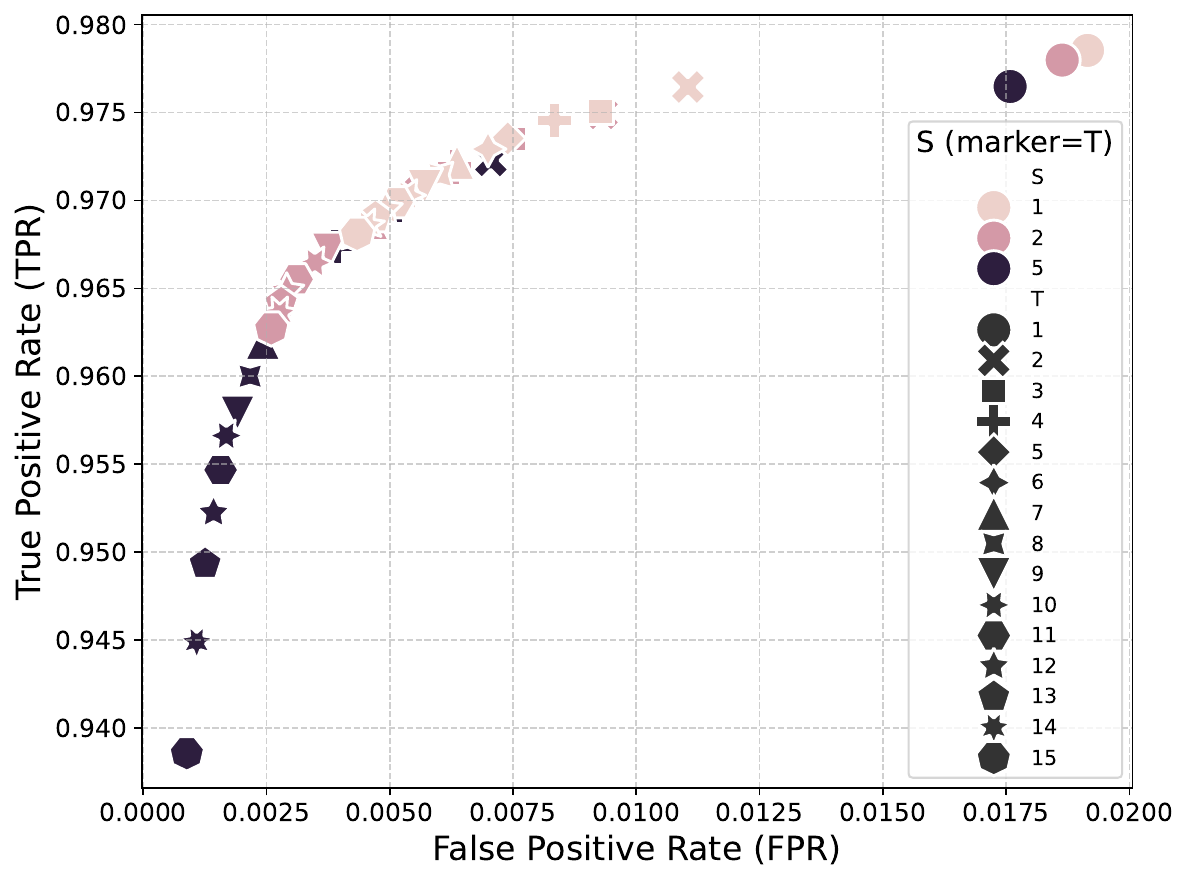}
  }
  \hfill
  \subfloat[TPR vs FPR (M=30)\label{fig:3}]{
    \includegraphics[width=0.3\textwidth]{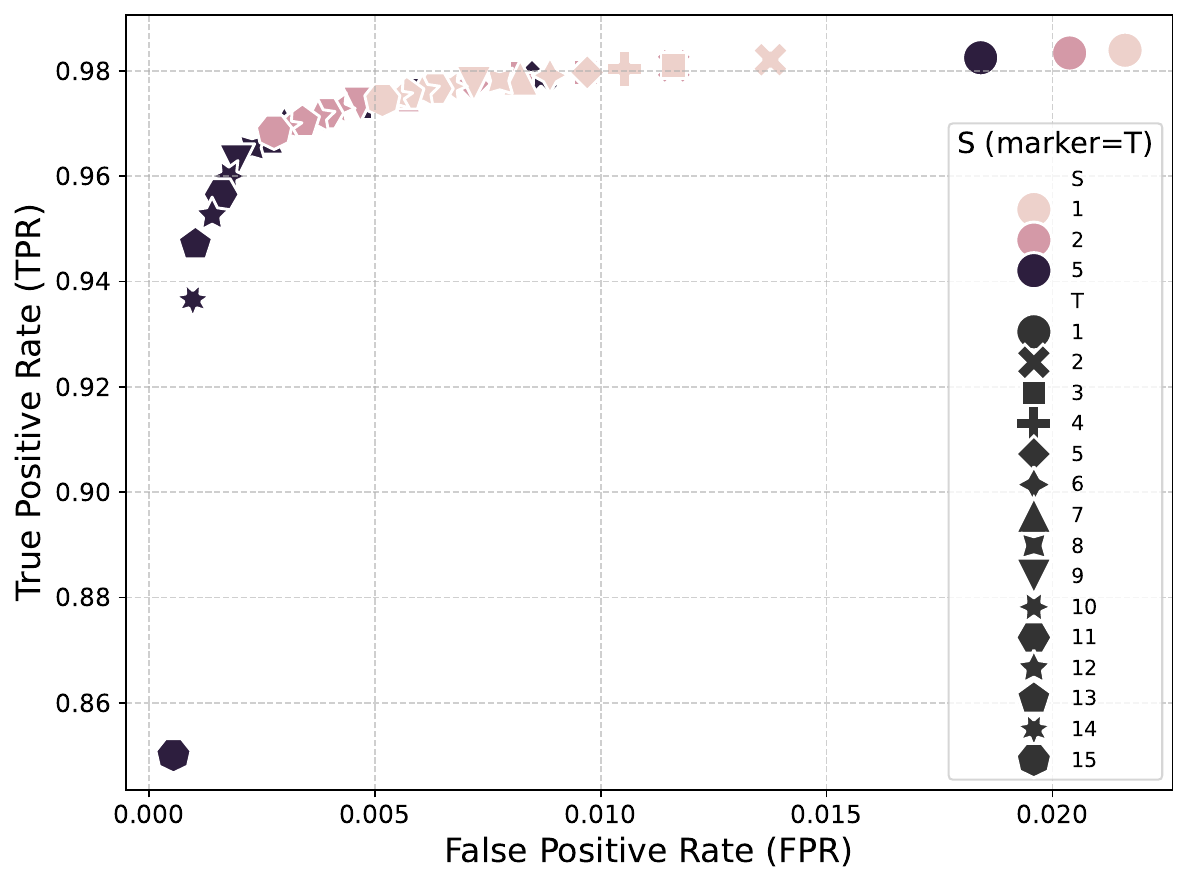}
  }
  \vspace{0.6em} 

  \subfloat[TPR vs FPR (M=40)\label{fig:5}]{
    \includegraphics[width=0.3\textwidth]{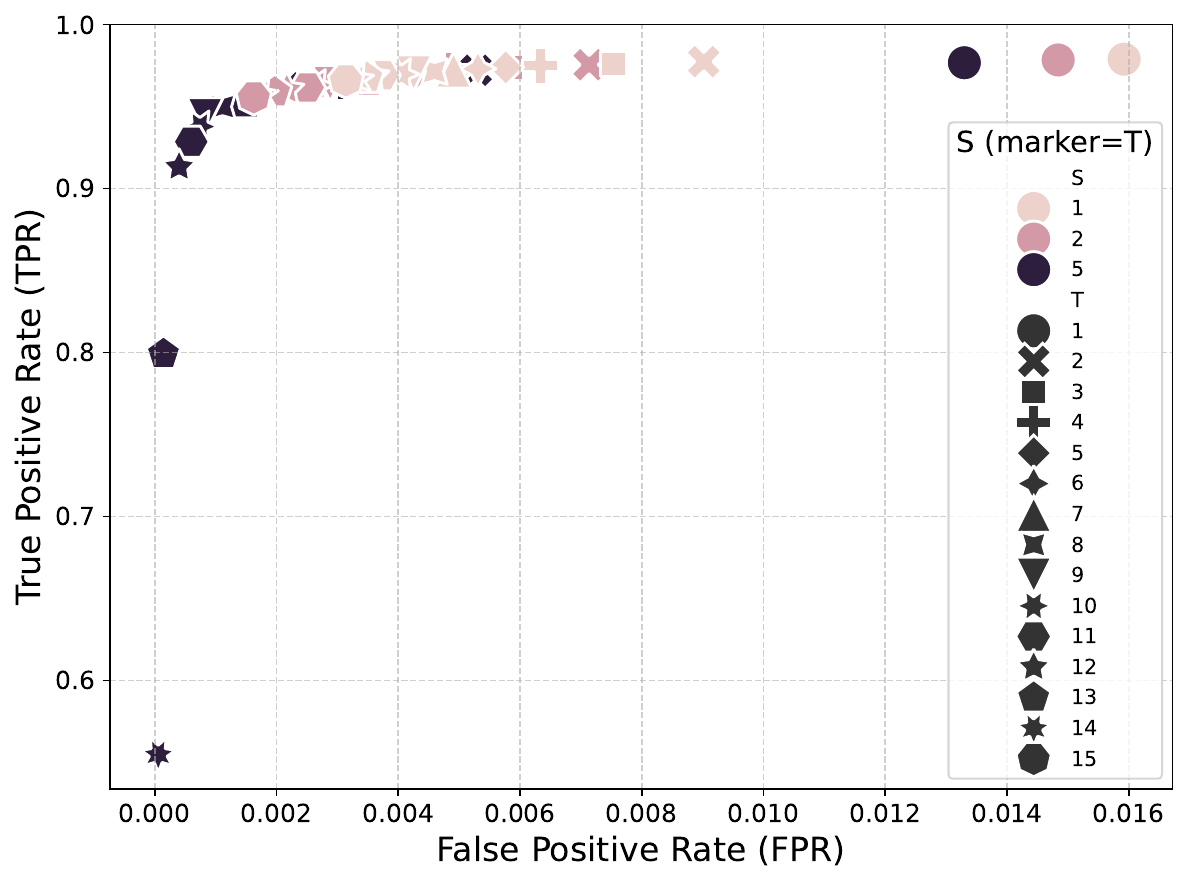}
  }
  \hspace{1cm}
  \subfloat[TPR vs FPR (M=50)\label{fig:6}]{
    \includegraphics[width=0.3\textwidth]{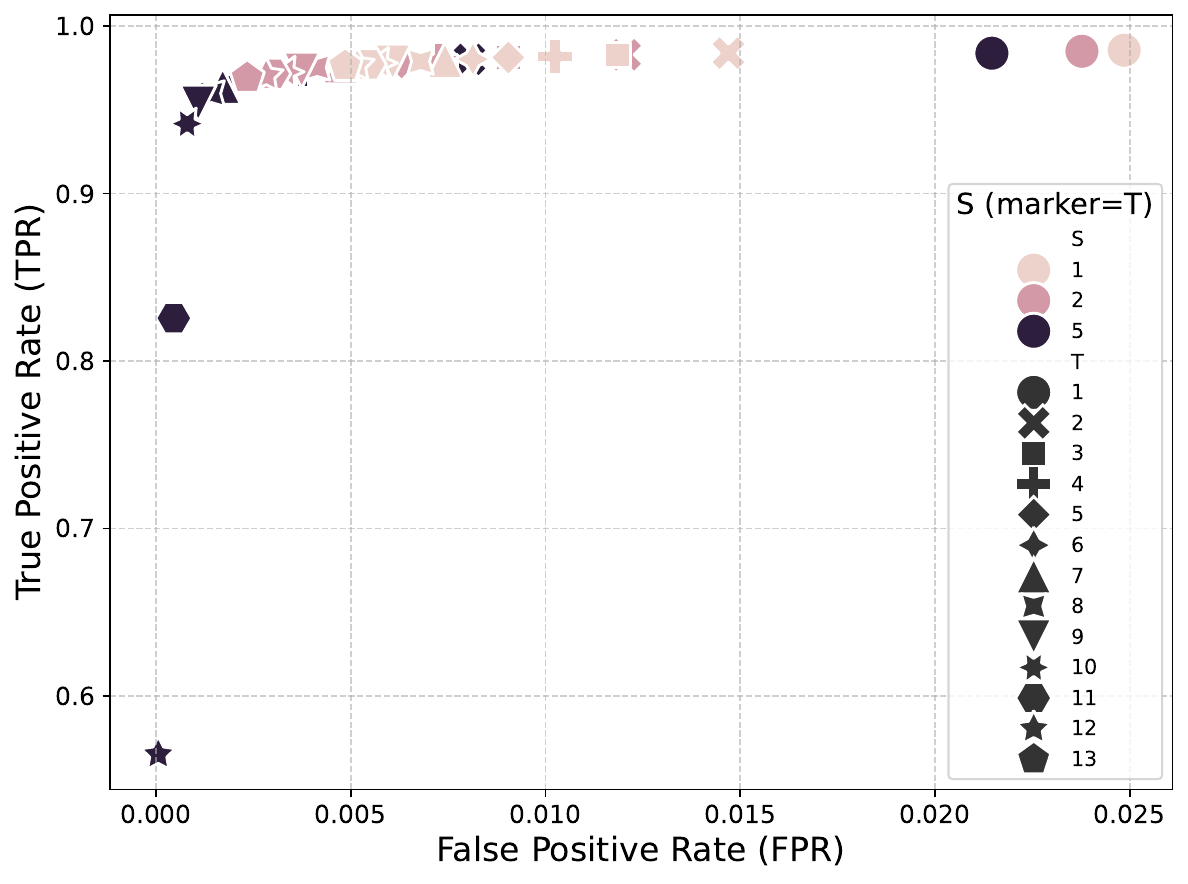}
  }

  \caption{True Positive Rate (TPR) vs. False Positive Rate (FPR) for different values of $M \in \{10, 20, 30, 40, 50\}$. Each subplot corresponds to a specific $M$ value, illustrating how the detection performance varies with the parameters $T$ and $S$. Colors represent different $S$ values, while marker styles indicate the corresponding $T$ values.}
  \label{fig:tpr_vs_fpr}
\end{figure*}

The proposed defense depends on several hyperparameters:
\begin{enumerate}
    \item a threshold-based voting mechanism, which classifies a sample as malicious if the number of masked versions labeled as malicious equals or exceeds the threshold;
    \item the mask size $M$, which determines the percentage of the file masked in each version;
    \item the stride size $S$, which specifies the number of bytes skipped between consecutive masked versions.
\end{enumerate}
As shown in Figure \ref{fig:tpr_vs_fpr}, the threshold determines the sensitivity of the detection system and plays a crucial role in balancing the trade-off between TPR and FPR. A lower, conservative threshold significantly increases the true positive rate at the cost of of increasing the false positive rate. In contrast, a higher threshold value significantly reduces the false positive rate but increases the risk of misclassifying malicious files. For example, the TPR of ByteShield-MalConv(M=20, S=1) decreases from 0.9785 to 0.9681 as the threshold value increases from 1 to 15. Conversely, its FPR decreases from 0.0192 to 0.0043.

Regarding the mask size $M$, the results indicate that there is relatively little variation in overall model performance, even though the actual proportion of occluded bytes significantly differs. Note that a mask size $M=10$ preserves most of the input content whereas a mask size $M=50$ occludes half of it. Despite this stark difference, the TPR and FPR metrics remain consistent across the different mask sizes. These findings align with the results presented in Table \ref{tab:masked_training}. 

With regard to the stride size $S$, empirical evidence shows that larger strides tend to have both lower TPR and FPR. For example, when fixing the mask size at $M=50$ and the threshold at $T=1$, the TPR gradually decreases from 0.98561 to $0.984984$ and $0.9838$ as the stride increases from $S=1$, to $S=2$, and $S=5$, respectively. Conversely, the FPR improves with larger strides, dropping from $0.0249$ to $0.0238$ and $0.0215$ over the same stride values.  Nevertheless, an edge case emerges with the configuration $M = 50$ and $S = 5$ when the threshold $T$ exceeds $10$. In this setting, the model experiences
a catastrophic drop in TPR and FPR, falling sharply to zero. This occurs because in this setting, only $10$ masked versions per input are generated during inference (see Table \ref{tab:number_masked_version_per_stride}).
However, when $T > 10$, the defense requires more votes than are available, making it mathematically impossible to classify any input as malicious. Consequently, the defense defaults to labeling all samples as benign, resulting in a complete collapse
of its detection performance.

\subsection{Comparison with Competing Methods (Clean Setting)}
\label{sec:clean_evaluation}
\begin{figure*}[ht]
    \includegraphics[width=\textwidth]{
    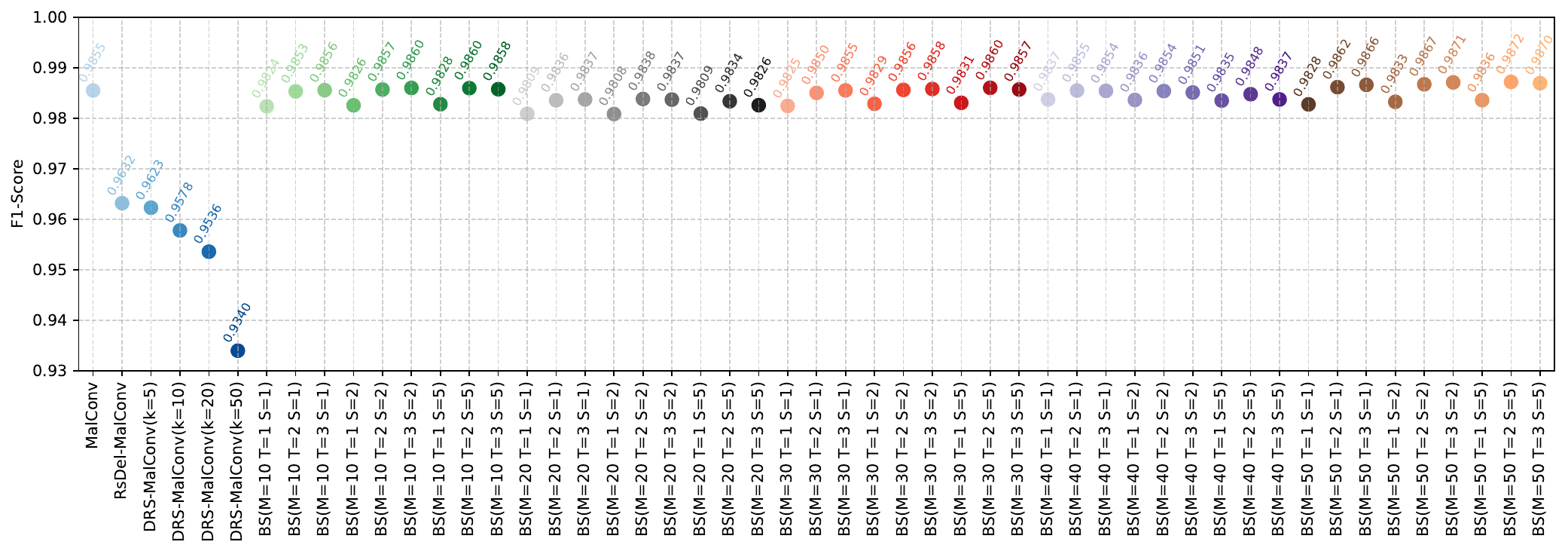}
    \centering
    \caption{F1-scores of all malware detectors across the full set of defense configurations.}
    \label{fig:scatter_plot_f1score_clean_performance}
\end{figure*}

\begin{figure*}[ht]
    \includegraphics[width=\textwidth]{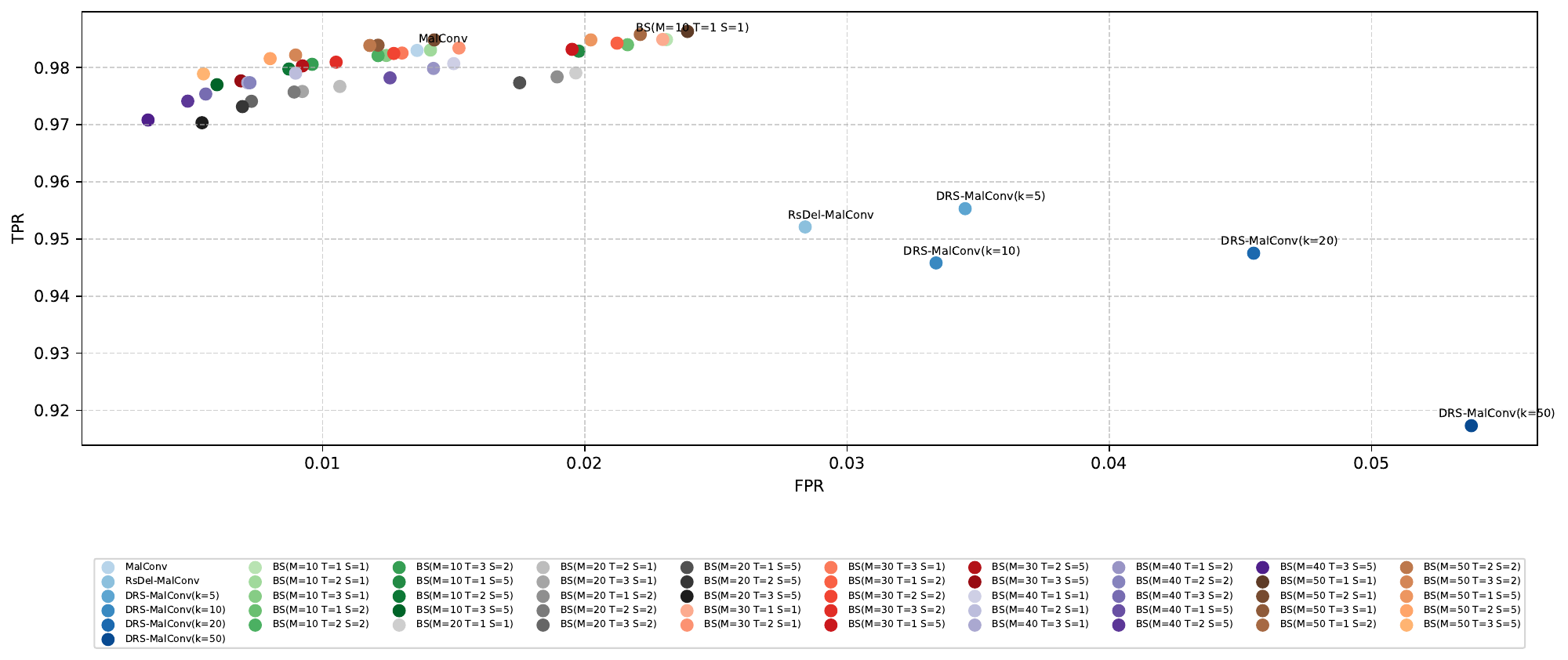}
    \centering
    \caption{True Positive Rate (TPR) versus False Positive Rate (FPR) for all malware detectors under the clean (non-attacked) setting.}
    \label{fig:scatter_plot_tpr_vs_fpr_clean_performance}
\end{figure*}

Next, we evaluate our defense alongside competing methods on clean examples, i.e., input files that have not been altered to evade detection. Figures \ref{fig:scatter_plot_f1score_clean_performance} and \ref{fig:scatter_plot_tpr_vs_fpr_clean_performance} present the F1-scores and the TPR–FPR trade-off of our approach and competing defenses on the test set.

Results demonstrate that our defense consistently outperforms competing defenses in terms of true positive rate, false positive rate, and f1-score. In contrast, defenses based on (de)randomized smoothing perform poorly, exhibiting FPR values above 0.025 and TPR values below 0.96.\footnote{We note that this reduction in performance for the DRS models is consistent with observations reported in their original papers \cite{10.1145/3605764.3623914,saha2024drsm}. This degradation arises from the fact that DRS operates on only a subset of the file’s bytes when performing classification.}
We also observe that configurations with larger values of $T$ generally deliver lower FPR. However,
models with larger $T$ values exhibit reduced robustness to evasion attacks. Consequently, an appropriate trade-off is achieved with moderate settings such as $T=2$ or $T=3$. \footnote{We include an extensive analysis of the robustness of our defense for different $T$ values in the Appendix.}

For clarity and brevity, in the next Sections we report only the results corresponding to $T=2$ with $S=1$ and $S=5$. This is because configurations with $T=1$ consistently yield lower F1-scores compared to those with $T > 1$. Furthermore, results for $T \in \{2,3,4,5\}$ and $S=1$ and $S=2$ are nearly identical, so including both would offer little additional insight. 

\subsection{Robustness Against Adversarial Attacks}
\label{sec:adversarial_evaluation}

We now assess the robustness of the malware detectors against adversarial attacks specifically designed to evade end-to-end malware detectors by injecting adversarial payloads in various locations within the PE files (Cf. Figure \ref{fig:pe_structure}). The attacks are the following:
\begin{enumerate}
    \item Padding attack \cite{8844597,8553214}, which injects the adversarial payload at the end of the executable file.
    \item Shift attack \cite{10.1145/3473039}, which injects the adversarial payload between the PE headers and the first section.
    \item Code caves attack \cite{YUSTE2022102643}, which splits the adversarial payload into multiple parts and injects them at the end of existing sections.
    \item Section injection attack \cite{9437194}, which splits the payload and injects the parts into newly-created sections. In our work, we limit the number of newly created sections to five.
\end{enumerate}

Due to the computationally intensive nature of generating adversarial payloads, we used
a reduced-size test set consisting of 500 randomly selected
malicious executables. This approach enabled us to obtain
valuable insights and results within a manageable timeframe.
It is worth noting that the size of our test set is comparable
to those used in prior work \cite{8844597,8553214,10.1145/3473039,YUSTE2022102643,9437194,saha2024drsm,287238,gibert2024certifiedadversarialrobustnessmachine,gibert2023_randomizedsmoothing}.

For each type of adversarial manipulation, we inject payloads of size 10\%, 20\%, 50\%, or 100\% of the original file. This ensures that the injected payload size remains proportional to the file’s original size and allows us to investigate the performance of the detectors under increasing perturbation sizes. Note that using a fixed perturbation size for every sample is not fair because the underlying file sizes vary widely. For example, a payload of 10,000 bytes represents a substantial modification for a small file (e.g., 50,000 bytes) but is almost negligible for a much larger file (e.g., 600,000 bytes). As a result, the level of adversarial manipulation becomes inconsistent across samples. The original implementation of the shift and padding attacks relied on FGSM \cite{DBLP:journals/corr/GoodfellowSS14} to craft the injected payload. This choice was reasonable given that early versions of these attacks added only a small number of bytes, i.e., 10,000 for the append attack and 4,096 for the shift attack. However, FGSM cannot be directly applied to randomized and (de)randomized smoothing defenses, and assumes white-box access, which is an unrealistic threat scenario. In practice, attackers rarely have access to the model gradients, and the stochasticity introduced by the aforementioned defenses further destabilizes gradient estimates. Therefore, in our work we optimize the payloads with Nevergrad's \cite{nevergrad} DoubleFastGADiscreteOnePlusOne optimizer, which avoids the limitations of FGSM and enables the generation of payloads of arbitrary size without requiring access to model gradients. To generate the payloads, we first extract bytes from benign executables and then optimize them using Nevergrad’s DoubleFastGADiscreteOnePlusOne optimizer with a budget of 3,000 objective evaluations.\footnote{We include an extended analysis in the Appendix where we compare the effects of initializing the adversarial payload with benign content and optimizing it with DoubleFastGADiscreteOnePlusOne. We will release the full implementation and model weights after acceptance.}

\begin{figure*}[htbp]
  \captionsetup[subfloat]{font=scriptsize} 

  \subfloat[Padding attack\label{fig:padding_results}]{
    \includegraphics[width=0.4\textwidth]{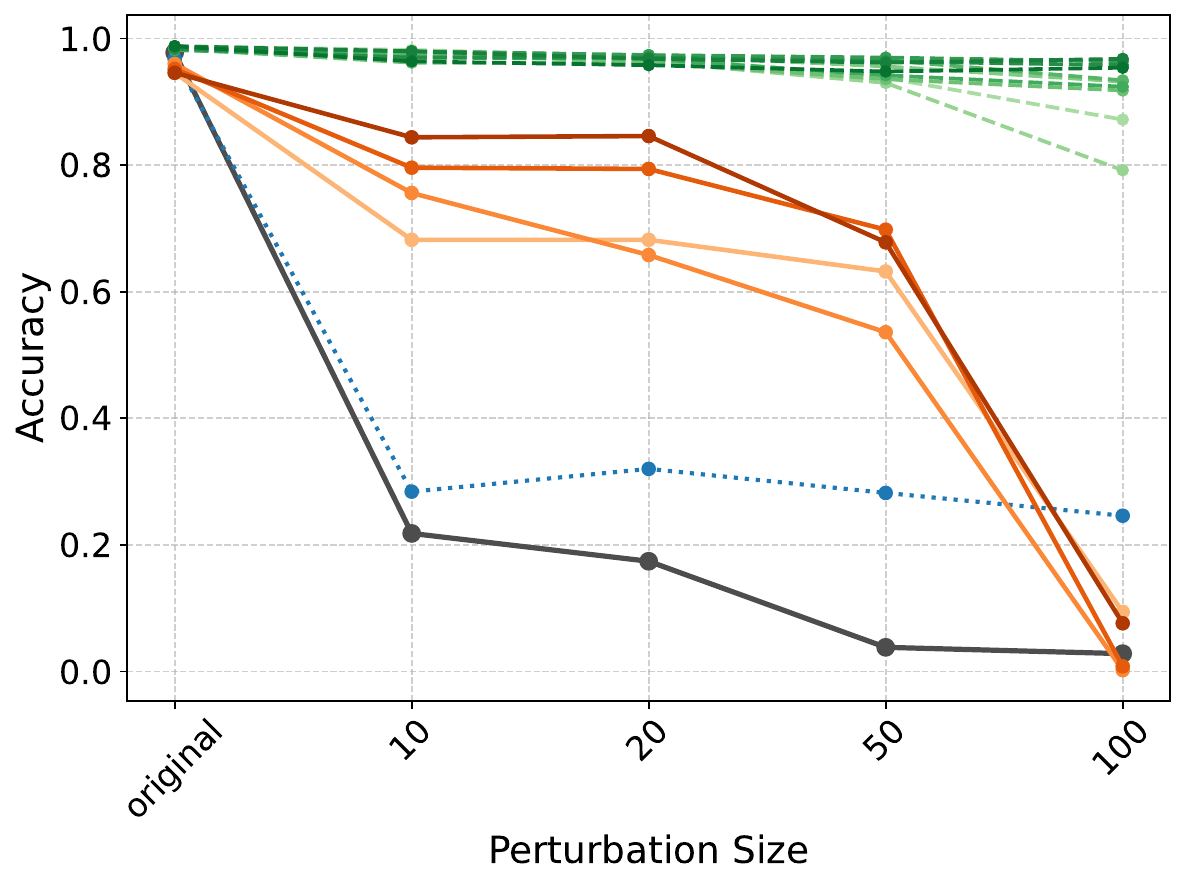}
  }
  \subfloat[Shift attack\label{fig:shift_results}]{
    \includegraphics[width=0.5\textwidth]{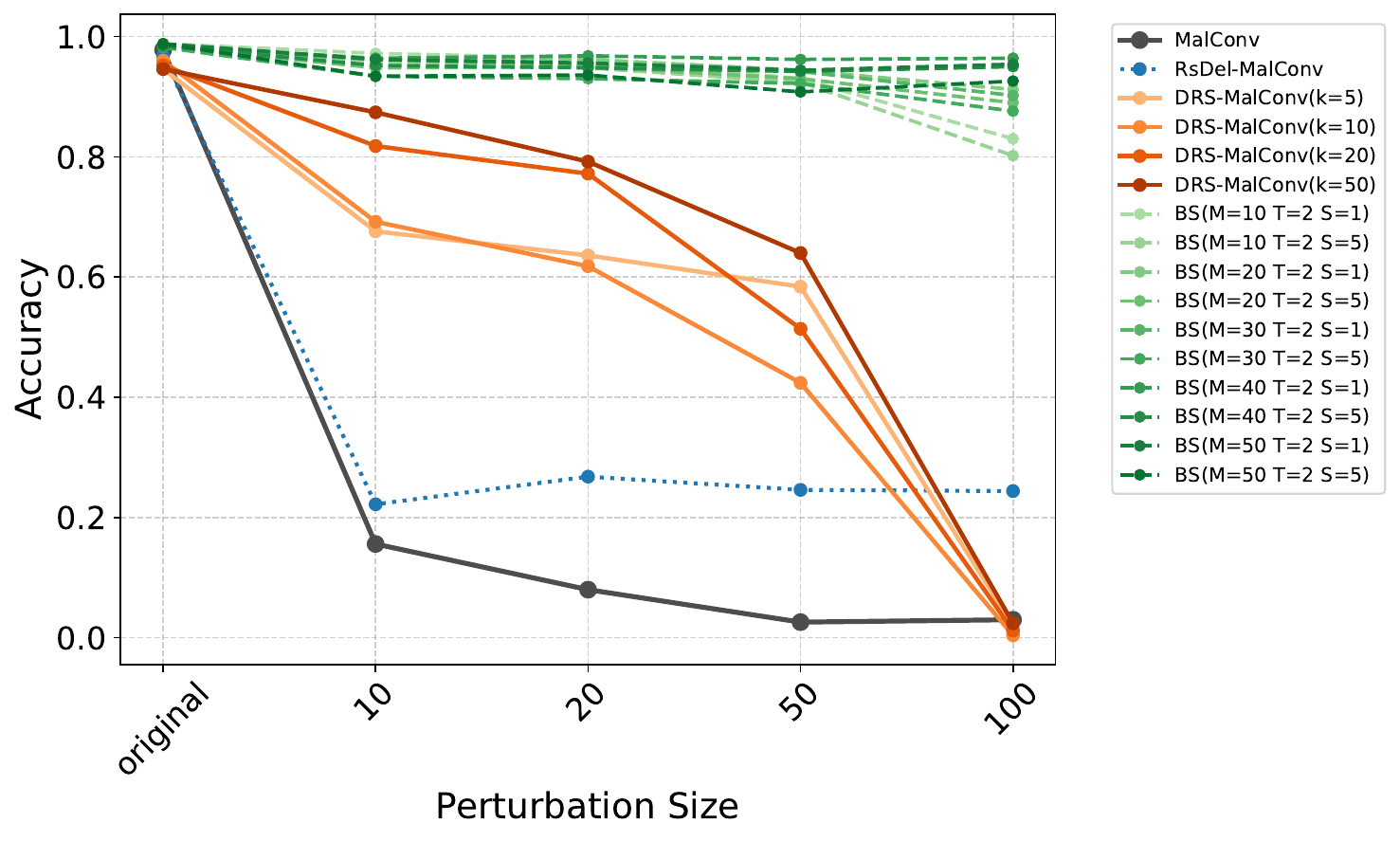}
  }\hfill

  \vspace{0.6em} 

  \subfloat[Code caves attack\label{fig:code_caves_results}]{
    \includegraphics[width=0.4\textwidth]{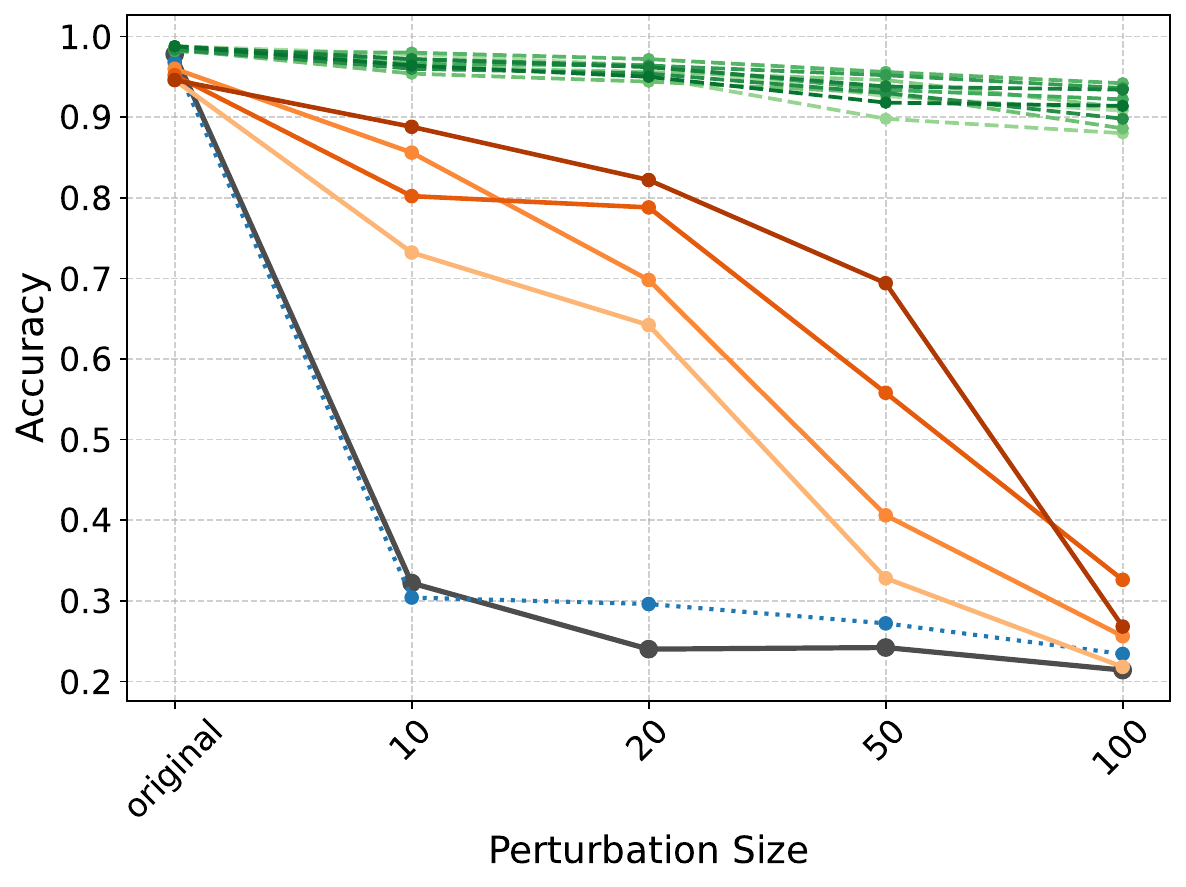}
  } 
  \subfloat[Section injection attack\label{fig:section_injection_results}]{
    \includegraphics[width=0.4\textwidth]{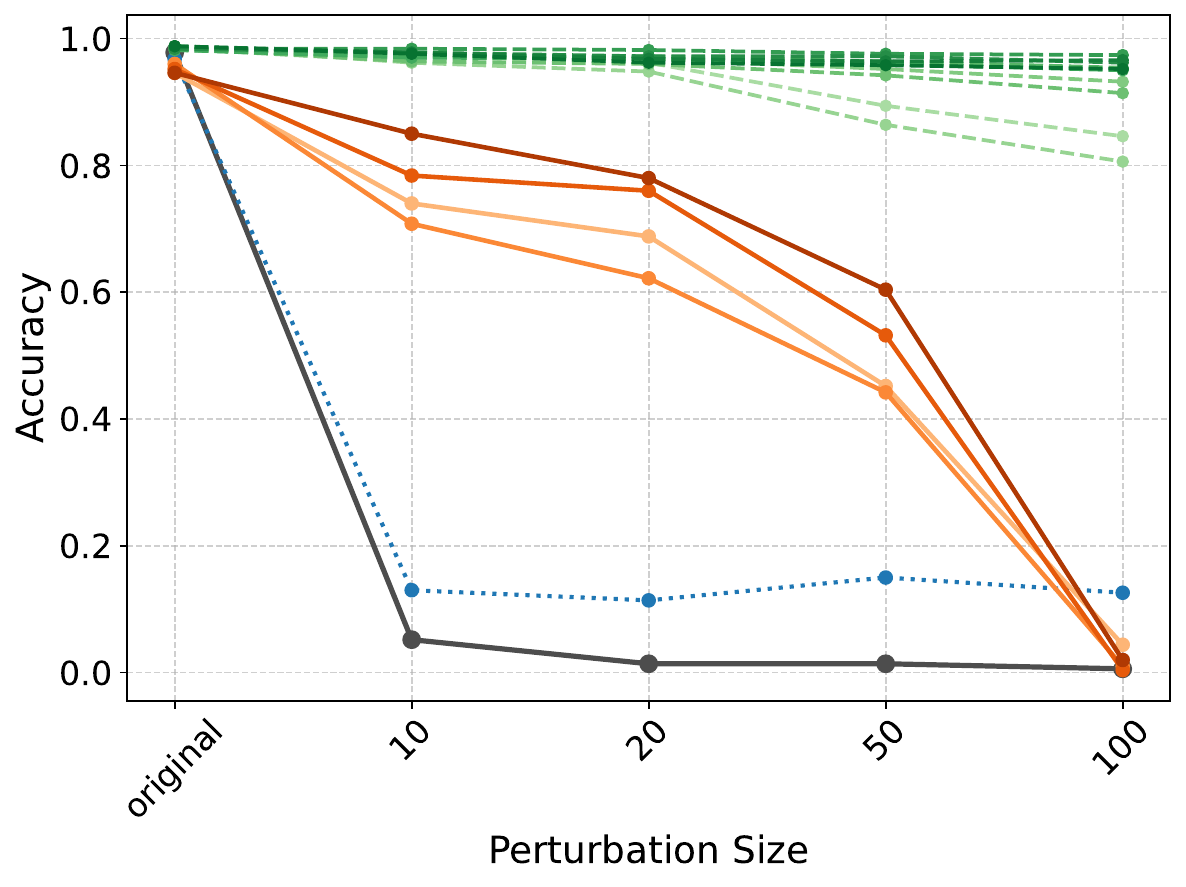}
  }
\caption{
MalConv-based malware detectors accuracy under increasing perturbation sizes (0–100\%).}
  \label{fig:adv_attackes_accuracy}
\end{figure*}

The results are shown in Figure~\ref{fig:adv_attackes_accuracy}. A clear trend emerges: the vanilla MalConv model experiences a dramatic drop in detection accuracy, falling from 98\% to around 20\% when the injected payload reaches just 10\% of the original file size. As the payload increases to 20\%, 50\%, and 100\%, the accuracy continues to deteriorate, eventually approaching zero. This vulnerability arises because vanilla MalConv is trained on benign and malicious samples without any defense mechanism. Our attacks inject content derived from benign files and then optimize it; as a result, the model encounters artifacts and byte patterns characteristic of benign data and is misled into classifying the adversarial sample as benign.

The second observable trend is the performance degradation of RsDel-MalConv, which relies on randomized deletions. For each input, the method produces $N$ noisy variants in which every byte is independently deleted with probability 97\%, an extremely aggressive deletion policy. This deletion policy becomes ineffective against attacks that inject adversarial payloads of even a few hundred bytes because it removes too much of the original malicious signal. The deletion process treats all bytes equally, erasing both the original and the injected content indiscriminately. As a result, the surviving 3\% of the file forms a highly distorted and noisy fragment that lacks the key discriminative features required for accurate detection.

The third observable trend is the drop in accuracy of the (de)randomized smoothing model, DRS-MalConv. In this approach, each executable is partitioned into $k$ chunks, which are classified independently before applying a majority vote to obtain the final prediction. DRS-MalConv achieves higher detection accuracy than both the baseline model and RsDel-MalConv when the injected payload occupies only 10\% or 20\% of the file, as the adversarial content affects only a minority of the chunks. However, its performance declines sharply as the payload grows, eventually reaching nearly 0\% accuracy when the injected content accounts for 100\% of the original size. At this point, the adversarial payload dominates the file, effectively overriding the majority-vote mechanism and causing the model to misclassify the adversarial sample as benign.

The final observable trend is the unmatched performance of our defense, ByteShield, particularly for configurations with larger mask sizes $M \in \{40, 50\}$. Unlike randomized ablation methods that remove individual bytes, ByteShield masks contiguous regions of the file across different noisy versions. Lower mask sizes $M \in \{10,20\}$ offer noticeably weaker robustness because the masked regions are simply too small to effectively occlude large adversarial payloads. In contrast, large mask sizes $M \in \{40, 50\}$ occlude a substantial portion of the input, increasing the probability that the adversarial payload is either fully or partially masked. Furthermore, ByteShield’s robustness against attacks that distribute the payload across multiple regions (e.g., code-caves and section-injection attacks) further demonstrates that even partial occlusion is sufficient to disrupt the adversarial signal. This is because not all regions containing the distributed adversarial payload contribute equally to the model's output. In these cases, masking the more influential regions significantly weakens the effectiveness of the attack. As a result, ByteShield consistently succeeds in labeling the adversarial examples as malware, even when the payload is dispersed throughout the file.

\subsection{Temporal Robustness}
\label{sec:temporal_robustness}
We use the BODMAS dataset to assess the temporal robustness of the malware detectors. The BODMAS dataset provides chronologically annotated malware and goodware samples collected from September 2019 to September 2020. Each model is trained on the samples from the EMBER dataset and then evaluated progressively on newer samples without retraining. For each month, we compute the F1-score and we summarize the degradation over time using the Area Under Time (AUT) metric \cite{235493}, which integrates the performance curve across all future evaluation points. Table \ref{tab:temporal_robustness} reports the F1 scores for each month and the AUT scores for the baseline detectors and ByteShield-MalConv with $M=50$, $T=2$ and $S=1$. Among the baselines, the vanilla MalConv shows modest resilience over time, whereas RsDel-MalConv, and the DRS-MalConv variants exhibit substantial degradation across multiple months. ByteShield-MalConv achieves the highest AUT score and consistently strong F1-scores across nearly all evaluation periods, demonstrating superior temporal robustness compared to all competing methods.\footnote{We omit the remaining ByteShield configurations, as Sections~\ref{sec:clean_evaluation} and~\ref{sec:adversarial_evaluation} showed that they achieve comparable performance on clean data while offering lower robustness against adversarial attacks. However, the omitted configurations exhibit similar temporal robustness to ByteShield-MalConv (M=50, T=2, S=1).}


\begin{table*}[ht]
\centering
\caption{Monthly F1-scores and AUT values for all baseline detectors and ByteShield-MalConv ($M=50$, $T=2$, $S=1$) evaluated on the BODMAS dataset from September 2019 to September 2020.}
\label{tab:temporal_robustness}
\resizebox{\textwidth}{!}{%
\begin{tabular}{l|ccccccccccccc|c}
\toprule
\textbf{Model}                    & \textbf{09/2019} & \textbf{10/2019} & \textbf{11/2019} & \textbf{12/2019} & \textbf{01/2020} & \textbf{02/2020} & \textbf{03/2020} & \textbf{04/2020} & \textbf{05/2020} & \textbf{06/2020} & \textbf{07/2020} & \textbf{08/2020} & \textbf{09/2020} &\textbf{AUT} \\ \midrule
MalConv                  & 0.9239          & 0.8958           & 0.9022          & 0.9264          & \textbf{0.9423} & \textbf{0.9122} & 0.9436          & 0.9141          & 0.9231          & 0.9057                      & 0.8890                               & 0.9138          & 0.8725          & 0.9160          \\
RsDel-MalConv            & 0.9100          & 0.8916           & 0.8701          & 0.8946          & 0.8956          & 0.8641          & 0.9278          & 0.9055          & 0.9149          & 0.8861                      & 0.8895                               & \textbf{0.9466} & \textbf{0.9362} & 0.8955          \\ \midrule
DRS-MalConv(K=5)         & 0.8764          & 0.8576           & 0.8287          & 0.8651          & 0.9002          & 0.8574          & 0.9100          & 0.8710          & 0.8741          & 0.8468                      & 0.8611                               & 0.8775          & 0.8226          & 0.8693          \\
DRS-MalConv(K=10)        & 0.8560          & 0.8329           & 0.8115          & 0.8446          & 0.8736          & 0.7814          & 0.8054          & 0.8112          & 0.8774          & 0.8043                      & 0.8213                               & 0.7607          & 0.7921          & 0.7818          \\
DRS-MalConv(K=20)        & 0.8610          & 0.8304           & 0.8045          & 0.8459          & 0.8996          & 0.8428          & 0.9079          & 0.8607          & 0.8759          & 0.8310                      & 0.8455                               & 0.8719          & 0.8283          & 0.8684          \\
DRS-MalConv(K=50)        & 0.8678          & 0.8465           & 0.8447          & 0.8982          & 0.8944          & 0.8512          & 0.9167          & 0.9062          & 0.9113          & 0.8973                      & 0.8723                               & 0.8343          & 0.7782          & 0.8463          \\ \midrule
ByteShield-MalConv & \textbf{0.9367} & \textbf{0.92791} & \textbf{0.9161} & \textbf{0.9305}          & 0.9358          & 0.9008          & \textbf{0.9564} & \textbf{0.9413}          & \textbf{0.9398}          & \textbf{0.9236} & \textbf{0.9370} & 0.9178          & 0.8713          & \textbf{0.9265}     
\\ \bottomrule
\end{tabular}%
}
\end{table*}

\subsection{Computational Time}
\label{sec:computational_time}
\begin{table}[ht]
\caption{End-to-end detectors average training time per epoch.}
\label{tab:training_time}
\centering
\begin{tabular}{l|c}
\toprule
\textbf{Detector} & \begin{tabular}{@{}c@{}}\textbf{Training Time}\\\textbf{(hours/epoch)}\end{tabular} \\ \midrule
MalConv        &     2:44:45                           \\
RsDel-MalConv  &     3:11:42                           \\
DRS-MalConv-5    &     0:43:28                         \\
DRS-MalConv-10    &     0:26:30                          \\
DRS-MalConv-20    &     0:19:04                          \\
DRS-MalConv-50    &     0:13:50                          \\ \midrule
ByteShield-MalConv (M=10)   &  3:11:46                            \\
ByteShield-MalConv (M=20)   &  3:12:05                            \\
ByteShield-MalConv (M=30)   &  3:13:16                            \\
ByteShield-MalConv (M=40)   &  3:14:20                            \\
ByteShield-MalConv (M=50)   &  3:14:37                            \\ \midrule
\end{tabular}%
\end{table}

The proposed byte masking strategy involves generating multiple masked versions by applying a mask to a location within the given input, independently classifying these masked versions, and aggregating the results to determine the final classification. While this masking process introduces some computational overhead, particularly during inference, its impact during training is minimal as each example is masked only once per batch. Masking a file is a lightweight operation, resulting in only a small additional cost per epoch. In fact, Table \ref{tab:training_time}, shows that training times (calculated on a machine with an NVIDIA 4090 GPU) remain comparable to approaches that take as input the entire raw binary sequences like MalConv, and RSDel-MalConv. In contrast, DRS-MalConv's training time per epoch is significantly faster due to its design: it processes small, fixed-sized chunks rather than entire files. Because only a single chunk is sampled and processed for each example in each epoch, smaller chunk sizes directly translate to shorted training epochs. 

At inference time, however, our masking-based approach is slightly slower than single-pass models such as MalConv as it requires generating and evaluating N masked versions per input. Nonetheless, the use of larger masks and strides helps mitigate the overhead by reducing the number of masked versions generated per example. Table~\ref{tab:inference_time} reports the average inference time per example, demonstrating that, despite this additional cost, the method remains practical for deployment.
In particular, when using $M=50$, $S=5$, ByteShield-MalConv achieves an average inference time of $0.0064$ seconds, compared to $0.0021$ seconds for the baseline MalConv. Despite this modest increase, our defense remains highly efficient, especially when contrasted with randomized smoothing approaches such as RsDel-MalConv, which require 0.19 seconds per example.

\begin{table}[ht]
\centering
\caption{End-to-end detectors average inference time.}
\label{tab:inference_time}
\begin{tabular}{l|c|c}
\toprule
 \textbf{Detector}              & \begin{tabular}{@{}c@{}}\textbf{Inference Time}\\\textbf{(seconds/example)}\end{tabular} & \textbf{N}\\\midrule
\textbf{MalConv}        &   \textbf{0.0021}         &         \textbf{1}            \\
RsDel-MalConv  &   0.1923         &         100          \\
DRS-MalConv-5  &   0.0022              &         5          \\ 
DRS-MalConv-10 &   0.0022              &         10          \\ 
DRS-MalConv-20 &   0.0022              &         20          \\ 
DRS-MalConv-50 &   0.0022              &         50          \\ \midrule
ByteShield-MalConv (M=10, S=1, T=2)   & 0.0389   & 90                         \\
ByteShield-MalConv (M=10, S=2, T=2)   & 0.0207   & 45                         \\
ByteShield-MalConv (M=10, S=5, T=2)   & 0.0097   & 18                         \\
ByteShield-MalConv (M=20, S=1, T=2)   & 0.0348   & 80                         \\
ByteShield-MalConv (M=20, S=2, T=2)   & 0.0185   & 40                         \\
ByteShield-MalConv (M=20, S=5, T=2)   & 0.0089   & 16                         \\
ByteShield-MalConv (M=30, S=1, T=2)   & 0.0308   & 70                         \\
ByteShield-MalConv (M=30, S=2, T=2)   & 0.0166   & 35                         \\
ByteShield-MalConv (M=30, S=5, T=2)   & 0.0081   & 14                         \\
ByteShield-MalConv (M=40, S=1, T=2)   & 0.0268   & 60                         \\
ByteShield-MalConv (M=40, S=2, T=2)   & 0.0146   & 30                         \\
ByteShield-MalConv (M=40, S=5, T=2)   & 0.0073   & 12                         \\
ByteShield-MalConv (M=50, S=1, T=2)   & 0.0226   & 50                         \\
ByteShield-MalConv (M=50, S=2, T=2)   & 0.0125   & 25                         \\
\textbf{ByteShield-MalConv (M=50, S=5, T=2)}   & \textbf{0.0064}   & \textbf{10}                         \\ \bottomrule
\end{tabular}%
\end{table}

\section{Conclusions}
\label{sec:conclusions}
In this paper we present a novel byte masking scheme
that operates by (1) generating multiple masked versions of
the input file, (2) independently classifying each version, (3)
and aggregating the predictions with a threshold-based voting
mechanism to produce the final classification. The proposed
method consistently outperforms state-of-the-art defenses against adversarial
attacks specifically developed to evade end-to-end malware
detectors, establishing a new benchmark in terms of
adversarial accuracy. The success of our defense is attributed
to two key components: (1) the masking strategy, and (2) the
conservative threshold-based voting mechanism:
\begin{enumerate}
    \item The use of masking during training enables more accurate classifications of partially occluded executables. At inference time, the masking strategy fully or partially occludes the adversarial payload,
    neutralizing its influence on the classifier’s prediction.
    This allows our defense to correctly classify the executables even in the presence of large adversarial payloads.
    \item The conservative threshold-based decision logic ensures
    that our defense errs on the side of caution when making
    assessments. This approach allows the system to avoid
    risky classifications and ensures that it only classifies an
    executable file as benign when there is unanimity (or
    almost unanimity) across all masked variants, minimizing the risk of misclassifications.
\end{enumerate}

Together, these components have positioned our defense as
the most effective and reliable tool for end-to-end malware
detection, particularly under adversarial conditions.

\subsection{Future Work}
This work has examined the robustness of end-to-end malware detectors when faced with adversarial attacks that inject payloads of hundred or thousands of bytes, and has proposed a robust, byte masking scheme to counter them. 
However, end-to-end detectors are only a subset of the broader spectrum of detection techniques based on machine learning proposed for detecting malware. Other methodologies, such as feature-based detectors~\cite{10.1145/2857705.2857713,2018arXiv180404637A,harang2020sorel20m,GIBERT2022117957} and grayscale image-based detectors~\cite{Nataraj11-372,DBLP:journals/virology/GibertMPV19}, have also been shown vulnerable to adversarial attacks, albeit of a different nature. These techniques fundamentally differ from end-to-end detectors, resulting in distinct attack vectors and requiring tailored defensive strategies.

\section*{Acknowledgments}
D.~Gibert was supported by grant RYC2023-043607-I funded by MICIU/AEI/10.13039/501100011033 and FSE+. F.~Many\`a was supported by grant PID2022-139835NB-C21 funded by MCIN/AEI/10.13039/501100011033 and by ERDF, EU. 
This research project was made possible through the access granted by the Galician Supercomputing Center (CESGA) to its supercomputing infrastructure. The supercomputer FinisTerrae III and its permanent data storage system have been funded by the NextGeneration EU 2021 Recovery, Transformation and Resilience Plan, ICT2021-006904, and also from the Pluriregional Operational Programme of Spain 2014-2020 of the European Regional Development Fund (ERDF), ICTS-2019-02-CESGA-3, and from the State Programme for the Promotion of Scientific and Technical Research of Excellence of the State Plan for Scientific and Technical Research and Innovation 2013-2016 State subprogramme for scientific and technical infrastructures and equipment of ERDF, CESG15-DE-3114.

\section*{Data and Code Availability}
The EMBER and BODMAS datasets are available to the public and the source code~\footnote{\url{https://github.com/danielgibert/byteshield-malware}} of our approach will be made available under a MIT License after the paper is accepted.

\bibliographystyle{plain}
\bibliography{refs.bib}

\appendix
\section*{Appendix}
\section*{Attacks Implementation}
In this work, we introduce two key modifications to previously developed functionality-preserving attacks designed to evade end-to-end malware detectors. First, instead of initializing the adversarial payload with zeros or random bytes, we initialize it with benign content, i.e., byte sequences extracted from a set of benign Windows PE files not used during training. Second, we replace the original optimization algorithm in secml-malware \cite{demetrio2021secmlmalware} with Nevergrad's DoubleFastGADiscreteOnePlusOne, a gradient-free optimizer capable of handling high-dimensional, discrete byte search spaces.

\begin{figure}[ht]
    \includegraphics[width=0.6\columnwidth]{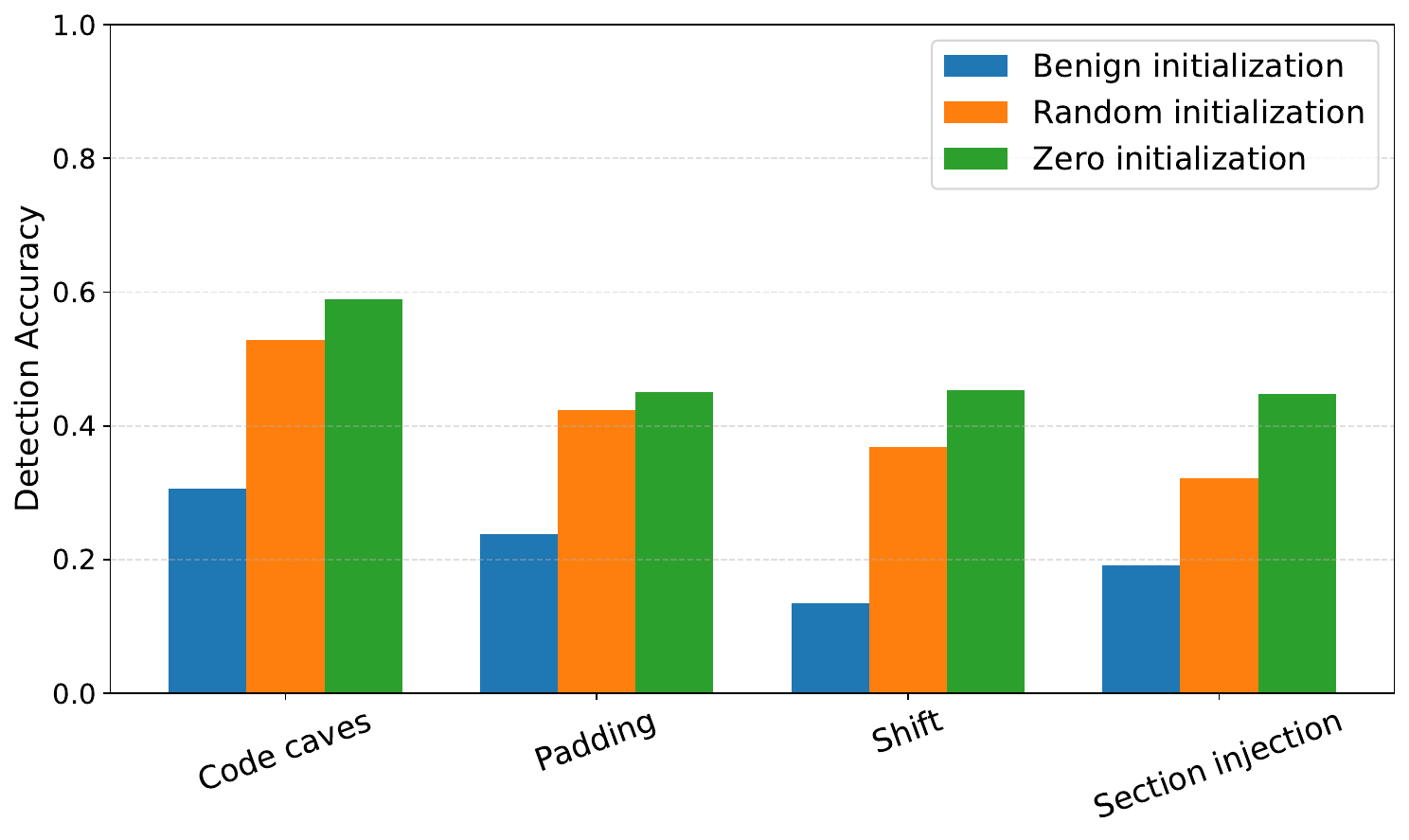}
    \centering
    \caption{Detection accuracy of MalConv against adversarial examples generated using the code-caves, padding, shift, and section-injection attacks with a perturbation budget of 10\% of the original file size and different payload initialization strategies.}
    \label{fig:genetic_algorithm_analysis}
\end{figure}

All attacks (padding, shift, code caves, and section injection) follow the authors' original implementation except for the modifications described above. The adversarial payload is always inserted according to the corresponding attack's manipulation strategy; however, it is initialized using benign bytes and then optimized with DoubleFastGADiscreteOnePlusOne to maximize evasion success. We experimented with several optimizers, but DoubleFastGADiscreteOnePlusOne consistently achieved the lowest detection rates. In Figure \ref{fig:genetic_algorithm_analysis}, we compare the detection accuracy of the MalConv detector against adversarial payloads generated with the code caves attack, the padding attack, the shift attack, and the section injection attack, each initialized with benign bytes, random bytes, or zero values. The results clearly indicate that initializing the payload with benign content substantially decreases the detection rate compared to the other initialization strategies indistinctively of the attack manipulation.

\section*{Robustness of ByteShield-MalConv for different configurations of T.}
\begin{figure*}[htbp]
  \centering
  \captionsetup[subfloat]{font=scriptsize} 

  \subfloat[Code caves attack\label{fig:code_caves_T_analysis}]{
    \includegraphics[width=0.42\textwidth]{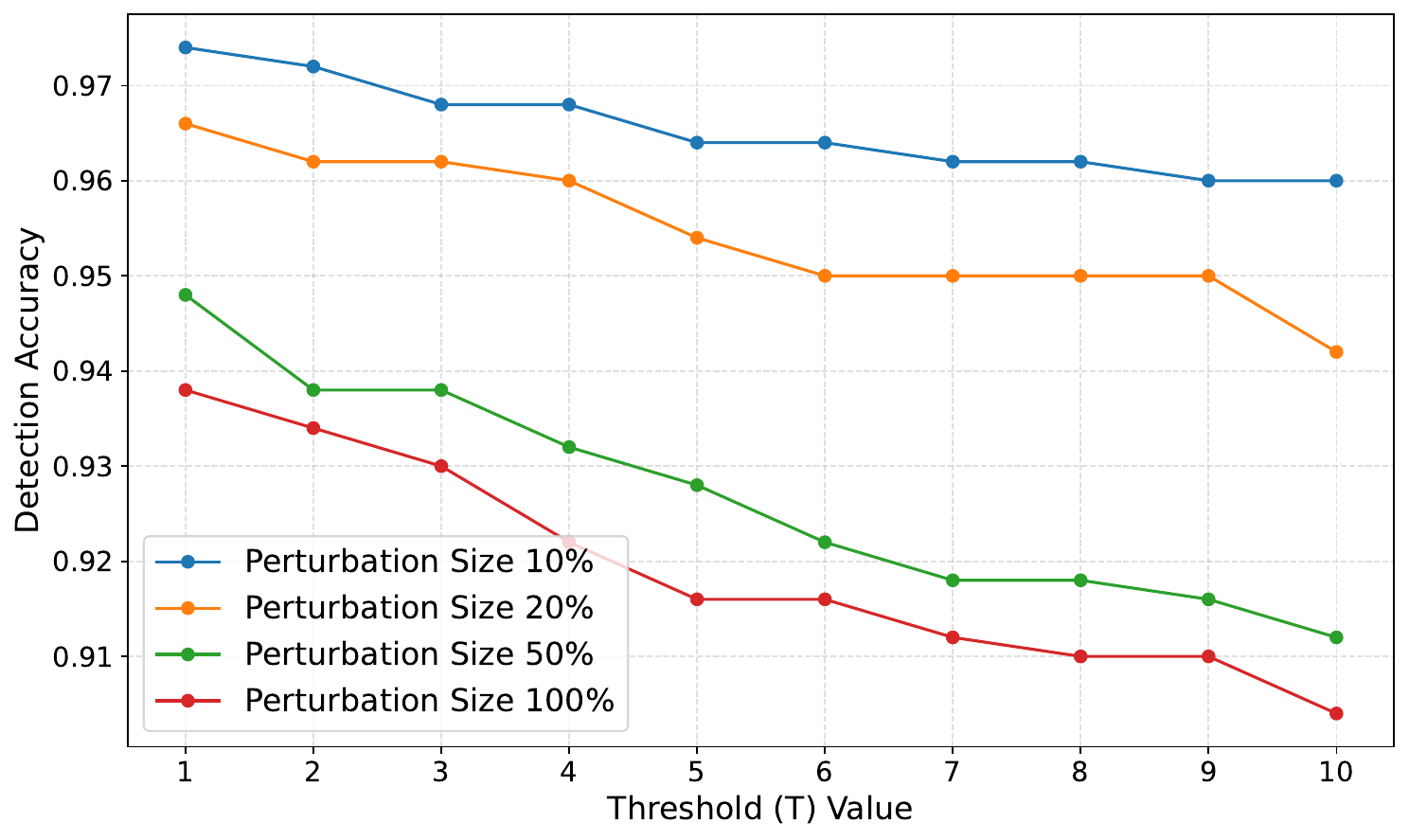}
  }\hfill
  \subfloat[Section injection attack\label{fig:section_injection_T_analysis}]{
    \includegraphics[width=0.42\textwidth]{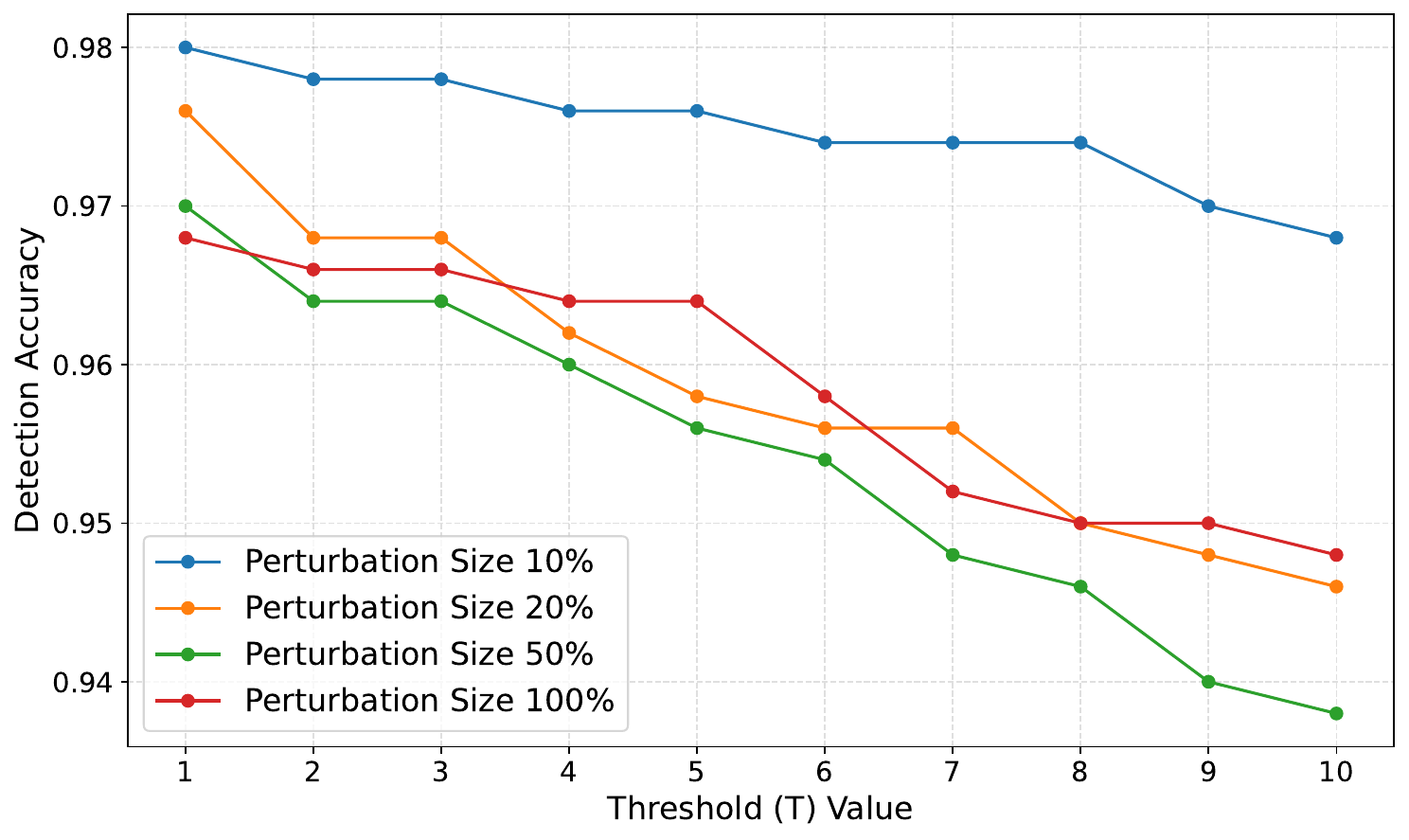}
  }\hfill
  \caption{Detection accuracy of ByteShield-MalConv ($M=50$) across threshold values $T\in\{1,2,3,4,5,6,7,8,9,10\}$ on adversarial examples generated using the code caves and section injection attacks with perturbation budgets of 10\%, 20\%, 50\%, and 100\%.}
  \label{fig:attacks_T_analysis}
\end{figure*}

In this Section, we evaluate the robustness of ByteShield-MalConv with $M=50$ and $S=1$ for different threshold values $T\in\{1,2,3,4,5,6,7,8,9,10\}$. Figures 
\ref{fig:code_caves_T_analysis}, and \ref{fig:section_injection_T_analysis} report the detection accuracy on adversarial examples crafted using the 
code caves and section injection attacks with perturbation budgets of 10\%, 20\%, 50\%, and 100\%, respectively.\footnote{We haven't included the Figures for the shift and padding attacks, as they exhibit the same trends as Figure \ref{fig:attacks_T_analysis}.} 

Across all attacks and perturbation budgets, a clear trend emerges: as the threshold $T$ increases, the detection accuracy on adversarial examples consistently decreases. Recall that $T$ determines how many masked versions must be classified "malicious" for our defense to label the input sample as malicious. Formally, a sample is classified as malicious if
$$num\_malicious \geq T$$
Larger values for $T$ require larger number of masked versions of the adversarial example to be classified as malicious. However, an adversarial payload can often flip the predictions of several masked versions preventing our defense from reaching the desired threshold. As a result, the attackers need to influence a smaller subset of the masked versions to bypass detection when $T$ is large.

\section*{Supplementary Experiments With Additional Architectures}
\begin{figure*}[htbp]
  \centering
  \captionsetup[subfloat]{font=scriptsize}

  \subfloat[Padding (AvastConv)]{
    \includegraphics[width=0.23\textwidth]{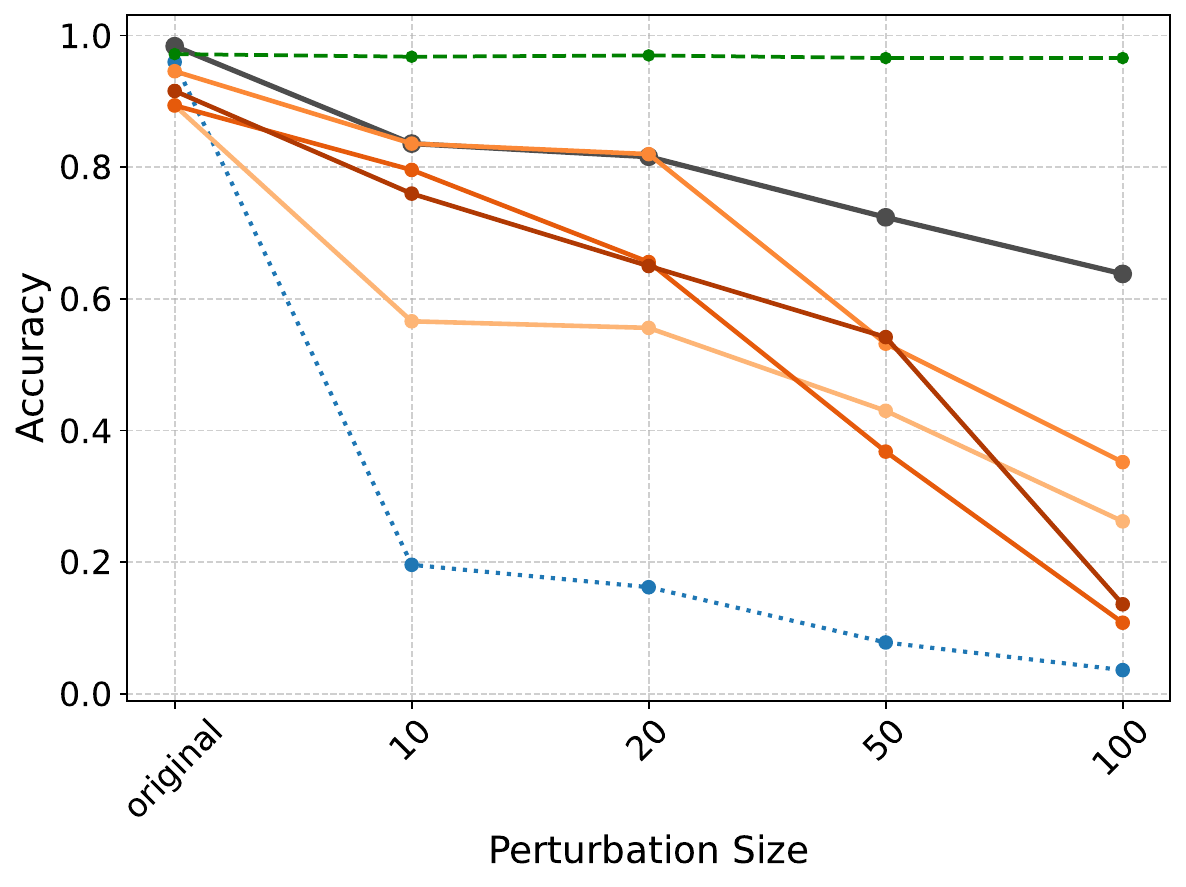}
  }\hfill
  \subfloat[Shift (AvastConv)]{
    \includegraphics[width=0.23\textwidth]{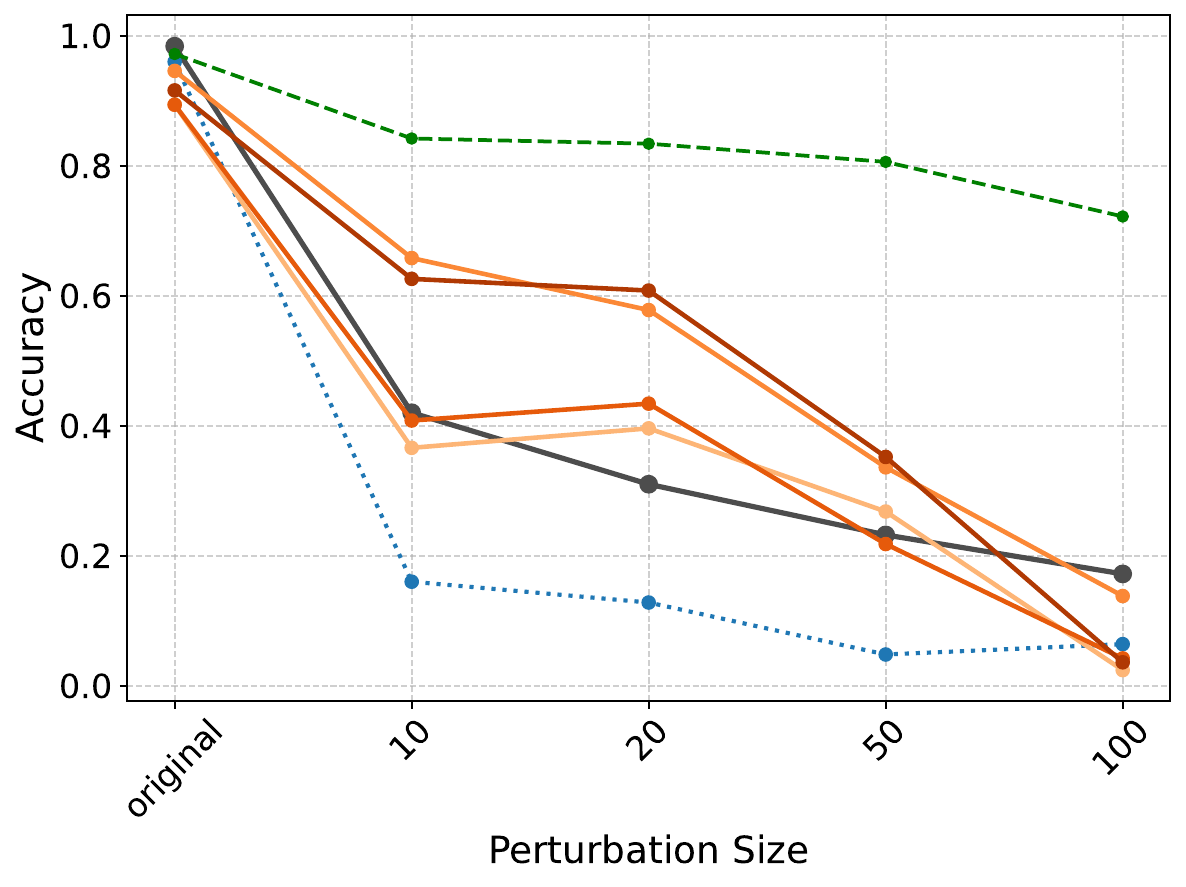}
  }\hfill
  \subfloat[Code caves (AvastConv)]{
    \includegraphics[width=0.23\textwidth]{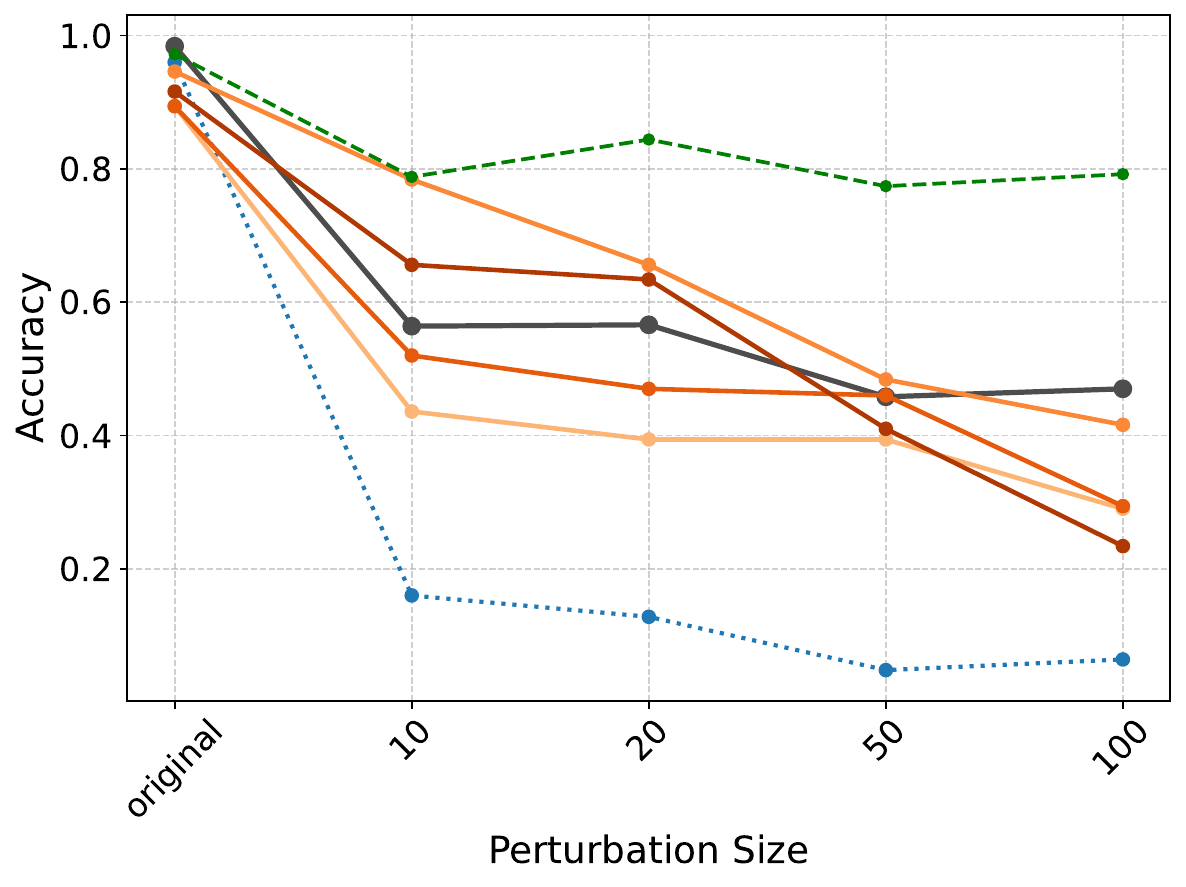}
  }\hfill
  \subfloat[Section injection (AvastConv)]{
    \includegraphics[width=0.23\textwidth]{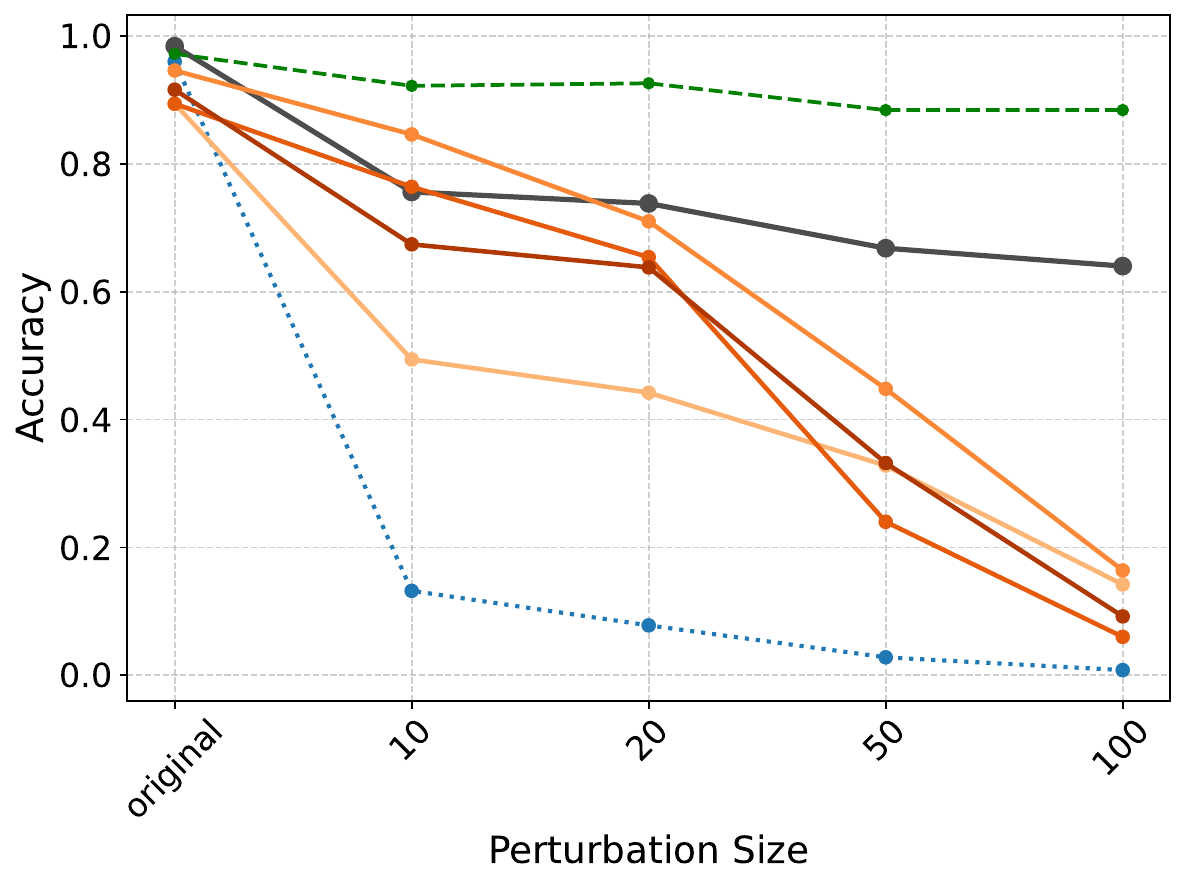}
  }

  \vspace{0.8em}

  \subfloat[Padding (NGramConv)]{
    \includegraphics[width=0.23\textwidth]{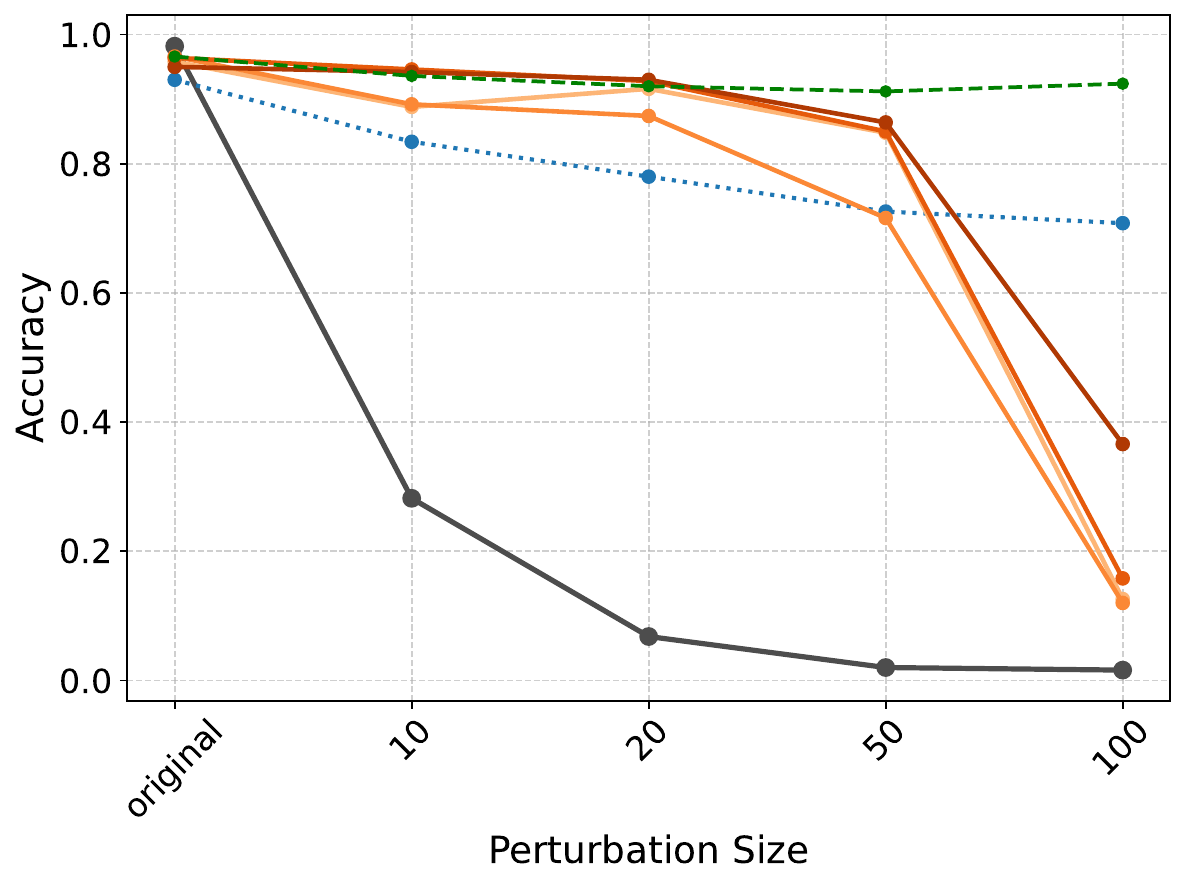}
  }\hfill
  \subfloat[Shift (NGramConv)]{
    \includegraphics[width=0.23\textwidth]{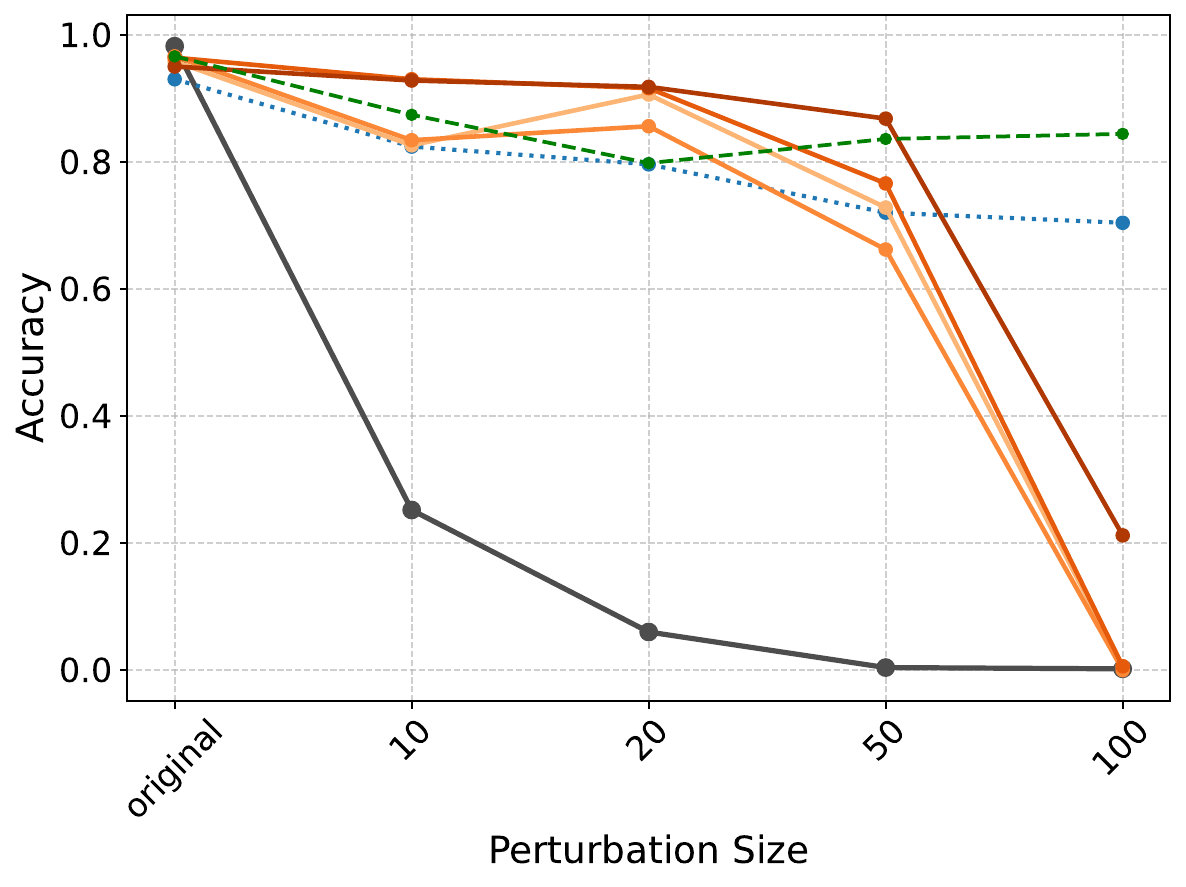}
  }\hfill
  \subfloat[Code caves (NGramConv)]{
    \includegraphics[width=0.23\textwidth]{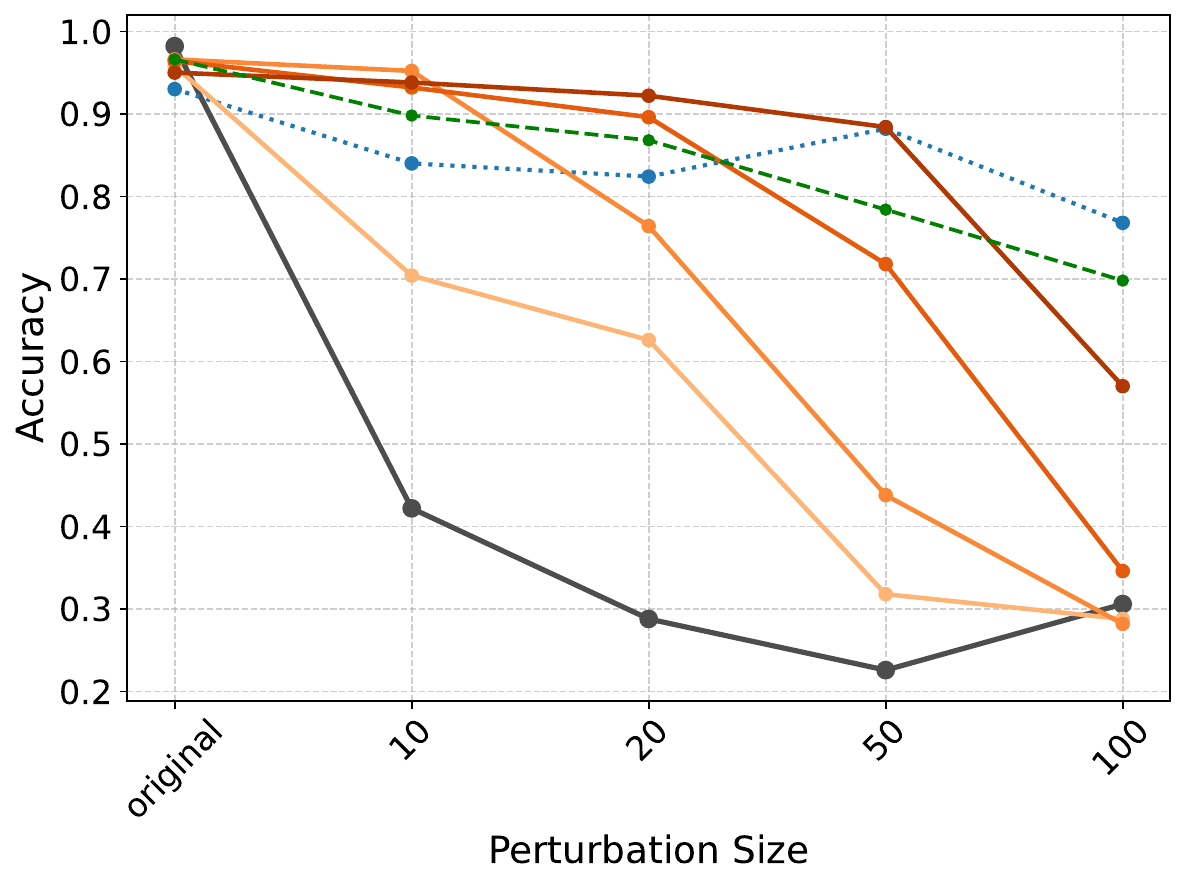}
  }\hfill
  \subfloat[Section injection (NGramConv)]{
    \includegraphics[width=0.23\textwidth]{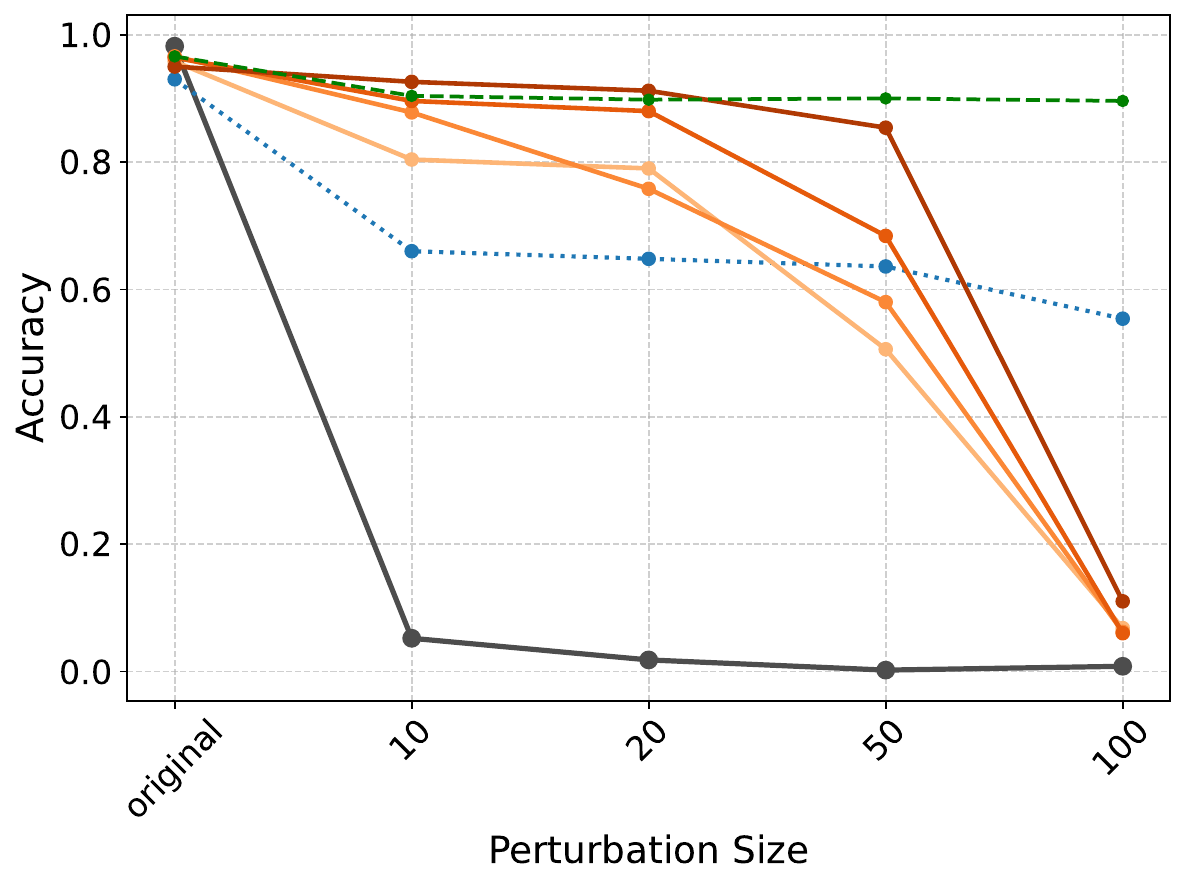}
  }

  \caption{
Accuracy degradation of AvastConv- and NGramConv-based malware detectors under increasing perturbation sizes (0–100\%).  
Legend: 
\textcolor{gray}{\large$\bullet$} Baseline \quad
\textcolor{blue}{\large$\bullet$} RSDel \quad
\textcolor{orange}{\large$\bullet$} DRS \quad
\textcolor{green}{\large$\bullet$} ByteShield.
}
\label{fig:adv_attack_accuracy_grid}
\end{figure*}

In this Section, we present additional results obtained using the AvastConv \cite{DBLP:conf/iclr/KrcalSBJ18} and NGramConv \cite{GIBERT2021102159} architectures to assess whether our findings generalize beyond MalConv. Figure \ref{fig:adv_attack_accuracy_grid} and  presents the detection accuracy of ByteShield-AvastConv and ByteShield-NGramConv with $M=50$, $S=1$ and $T=2$.

Similar to MalConv’s results, our defense consistently outperforms competing approaches when using AvastConv and NGramConv as backbones, particularly at larger perturbation sizes. We observe that the vanilla AvastConv model attains higher accuracy than RsDel-AvastConv. We hypothesize that these differences arise from the model's architectural characteristics, particularly the kernel sizes and depth of their convolutional layers. Models with larger kernels (e.g., MalConv) and deeper convolutional stacks (e.g., AvastConv), rely on wide receptive fields that become unstable under RsDel’s aggressive deletion, which disrupts long-range structure. In contrast, architectures using smaller kernels, such as NGramConv, operate on more local patterns and might still be able to learn deletion-invariant features.

\end{document}